\documentclass[twocolumn,letterpaper,10pt,groupedaddress,nofootinbib,aps,prd]{revtex4-2} % reprint
\pdfoutput=1
\usepackage[utf8]{inputenc}
\usepackage[colorlinks,citecolor=blue]{hyperref}
\usepackage{braket}
\usepackage{upgreek}
\usepackage[normalem]{ulem}
\usepackage[dvips]{graphicx}
\usepackage{hhline}
\usepackage{epsfig,amsmath,amssymb,verbatim,mathrsfs,array,layout,textcomp,latexsym,slashed,booktabs,color,mathtools}
\usepackage[caption=false]{subfig}
\usepackage{multirow,units,halloweenmath}
\usepackage{microtype} % for bad boxes
\usepackage{capt-of} % for non-float captions

\makeatletter

\renewcommand\onecolumngrid{%
\do@columngrid{one}{\@ne}%
\def\set@footnotewidth{\onecolumngrid}%
\def\footnoterule{\kern-6pt\hrule width 1.5in\kern6pt}%
}

\renewcommand\twocolumngrid{%
\def\footnoterule{%
\dimen@\skip\footins\divide\dimen@\thr@@
\kern-\dimen@\hrule width 0.5in\kern\dimen@}
\do@columngrid{mlt}{\tw@}
}

\makeatother
\newcommand{\tocless}[1]{%
\let\oldaddcontentsline\addcontentsline% store \addcontentsline
\renewcommand{\addcontentsline}[3]{}% make \addcontentsline a no-op
#1%
\let\addcontentsline\oldaddcontentsline% restore \addcontentsline
}

\newcommand{\Tr}{\operatorname{Tr}}

\allowdisplaybreaks

\begin{document}

\title{General Approach to Neutrino Mass Mechanisms with Sterile Neutrinos}

\author{Yale Fan}
\email{yalefan@gmail.com}
\affiliation{Department of Physics, University of Texas at Austin, Austin, TX 78712, USA}
\affiliation{Sandia National Laboratories, Albuquerque, NM 87185, USA}

\author{Anil Thapa}
\email{wtd8kz@virginia.edu}
\affiliation{Department of Physics, University of Virginia, Charlottesville, VA 22904, USA}

\begin{abstract}
We present a mathematical framework for constructing the most general neutrino mass matrices that yield the observed spectrum of light active neutrino masses in conjunction with arbitrarily many heavy sterile neutrinos, without the need to assume a hierarchy between Dirac and Majorana mass terms. The seesaw mechanism is a byproduct of the formalism, along with many other possibilities for generating tiny neutrino masses. We comment on phenomenological applications of this approach, in particular deriving a mechanism to address the long-standing $(g-2)_\mu$ anomaly in the context of the left-right symmetric model.
\end{abstract}

\maketitle

\tocless{\section{Introduction}}

Understanding the origin of the tiny neutrino masses required to explain the observed oscillation data \cite{ParticleDataGroup:2022pth} is a fundamental problem in particle physics. The simplest extension of the Standard Model (SM) that can accommodate such masses is arguably the seesaw mechanism \cite{Minkowski:1977sc, Yanagida:1979as, Gell-Mann:1979vob, Glashow:1979nm, Mohapatra:1979ia, Schechter:1980gr, Schechter:1981cv, Foot:1988aq}, wherein light neutrino masses are generated through a higher-dimension operator \cite{Weinberg:1979sa} suppressed by the scale of new physics $\Lambda$. To obtain the observed neutrino masses ($M_\nu \sim 0.1$ eV), $\Lambda$ is typically taken to be around $10^{14}$ GeV. However, the masses of the new degrees of freedom may depend parametrically on a combination of Yukawa couplings $y$ and $\Lambda$. For instance, in the type I seesaw mechanism, the Majorana neutrino mass reads as $y^2 v^2/\Lambda$, where $v$ is the electroweak symmetry breaking vacuum expectation value (VEV). Hence the masses of heavy right-handed sterile neutrinos can lie near the TeV scale if the Dirac mass of order $yv$ is near the keV scale.

In this paper, we propose a systematic method to construct neutrino mass matrices that reproduce the observed spectrum of light neutrino masses while incorporating any number of comparatively heavy sterile neutrinos, \emph{without} assuming a seesaw hierarchy between Dirac and Majorana masses. This approach opens new avenues for the physics involved in neutrino mass generation to be probed directly in experiments. Notably, the resulting non-seesaw configurations offer alternative scenarios in which the masses of sterile neutrinos lie at relatively low scales, making them potentially detectable even with $\mathcal{O}(1)$ Dirac Yukawa couplings.

The key observation is simple: in the limit of a large hierarchy between active and sterile neutrino masses, any viable mass matrix can be obtained by perturbing a texture that gives rise to a number of exactly massless neutrinos equal to the number of active neutrinos. We derive conditions that characterize all such ``seed'' textures. These conditions encompass and generalize all seesaw mechanisms that extend the SM by sterile neutrinos.

Within the seesaw paradigm, there exist many specific models for the entries of the Dirac and Majorana mass matrices. For instance, there may exist hierarchies \emph{among} the heavy Majorana masses for right-handed neutrinos, in the form of ``sequential dominance'' \cite{King:2003jb}. Our method, which encapsulates all possible scenarios with three light active neutrinos, both reproduces and generalizes these previously studied matrix textures. We take a bottom-up approach and do not discuss particular embeddings of these models into grand unified theories. Likewise, we remain agnostic as to whether the entries of the mass matrix are generated at tree level or loop level.

While prior investigations have used perturbation theory to understand or engineer neutrino mass hierarchies, they have so far been limited to the seesaw framework. These include analyses of seesaw-type textures in which the right-handed Majorana mass matrix does not have full rank \cite{Lindner:2001hr}, as well as studies of such textures using matrix analysis \cite{Besnard:2016tcs, Flieger:2020lbg}. Importantly, Kersten and Smirnov \cite{Kersten:2007vk} have observed in the setting of the type I seesaw that if a neutrino mass matrix is perturbatively close to a texture that results in massless light neutrinos, then the seesaw mechanism itself plays a negligible role in explaining the smallness of neutrino masses. The premise of our work, which forgoes seesaw assumptions altogether, is that perturbation theory provides a far more general approach to the problem of generating tiny neutrino masses than previously considered.

\tocless{\section{Framework}} \label{sec:framework}

The neutrino mass matrix $M$ appears in the Lagrang\-ian through a term $\nu_f^T M\nu_f$, where $\nu_f$ is a vector of flavor eigenstates.  As such, $M$ defines a bilinear rather than a sesquilinear form on flavor space.  A change of basis cor\-responds to a congruence transformation $M\mapsto U^T MU$ rather than a similarity transformation $M\mapsto U^{-1}MU$, where $U$ is a unitary matrix.  This distinction clarifies how physical neutrino masses are obtained by ``diagonalizing'' $M$, with the physical masses corresponding to the singular values rather than the eigenvalues of $M$ \cite{Haber:2020wco}.

Let $m$ and $n$ denote the number of active and sterile neutrinos, respectively.  For phenomenological applications, we set $m = 3$ while leaving $n$ arbitrary.  The singular values and eigenvalues of a matrix vary continuously with its entries, so as the $m$ smallest singular values of the neutrino mass matrix $M$ become arbitrarily small relative to the $n$ largest singular values (corresponding to $m$ active neutrinos), the entries of $M$ must approach a texture that has $m$ vanishing singular values.  Inverting this logic, we find the most general conditions under which a neutrino mass matrix $M$ in the flavor basis has $m$ vanishing singular values.  Any viable model of small neutrino masses can then be constructed as a perturbation of a solution to these general vanishing conditions.

Consider a complex symmetric mass matrix of the form
\begin{equation}
M = \left[\begin{array}{cc} \mathbf{0}_{m\times m} & B \\ B^T & D \end{array}\right]
\label{massmatrix}
\end{equation}
in the flavor basis, where $B$ is an $m\times n$ complex matrix, $D$ is an $n\times n$ complex symmetric matrix, and $N \linebreak[1] = \linebreak[1] m \linebreak[1] + \linebreak[1] n$.  We focus on perturbations of $M$ that keep the upper-left block of left-handed Majorana masses exactly zero, which amount to perturbations of $B$ and $D$.  We assume the perturbations are small relative to the nonzero entries of $B$ and $D$, thus preserving the hierarchical nature of the problem.

Since the rank of a matrix is the number of nonvanishing singular values, we can easily formulate a general method to obtain textures that lead to $m$ light neutrinos and $n$ heavy neutrinos.  Start with a matrix $M$ that adheres to certain physical constraints (e.g., symmetric and with a vanishing upper-left block as in Eq.\ \eqref{massmatrix}) and whose rank equals the number of sterile neutrinos $n$ (or equivalently, whose nullity equals the number of active neutrinos $m$).  Next, perturb this matrix in a way that respects the physical constraints.  Strictly speaking, this prescription results in $n$ light neutrino masses if $n < m$ and $m$ light neutrino masses if $n\geq m$.  Prior to the perturbation, $M$ has $m$ vanishing singular values.  After the perturbation, there remains a vanishing $m\times m$ block, so $M$ generically has $m - n$ vanishing singular values, leading to $n$ small but nonzero singular values.\footnote{First-order perturbation theory tells us that the resulting light neutrino masses are generically linear in the perturbations. In the special case of the seesaw, where $B\ll D$ and $B$ itself (more precisely, $\left[\begin{smallmatrix} 0 & B \\ \smash{B^\text{\scalebox{0.75}{$T$}}} & 0 \end{smallmatrix}\right]$) is the additive perturbation to $\left[\begin{smallmatrix} 0 & 0 \\ 0 & D \end{smallmatrix}\right]$, the light masses are instead quadratic in the perturbations because the first-order correction vanishes. \label{firstorder}}

The key question is then: how can we systematically construct inputs to this method, i.e., textures with $m$ vanishing singular values?  First consider the \emph{a priori} unrelated problem of constructing textures with $m$ vanishing eigenvalues.  Without loss of generality, we choose a basis where $D$ is diagonal with real entries $d_i\geq 0$, according to the Autonne-Takagi factorization of $D$ \cite{Autonne, Takagi}.  Then $M$ has $m$ vanishing eigenvalues if and only if
\begin{equation}
\begin{aligned}
\sum_{r = \lceil\frac{n - d}{2}\rceil}^{n - d} (-1)^r\sum_{|\alpha| = r} e_{2r - n + d}(d_{\alpha(1)}, \ldots, d_{\alpha(r)})& \\[-10pt]
{}\times\det(G(\alpha|\alpha)) &= 0
\end{aligned}
\label{conditions}
\end{equation}
for all $d = 1, \ldots, n$, where $\alpha$ is a strictly increasing integer sequence of length $r$ chosen from $1, \linebreak[1] \ldots, \linebreak[1] n$; $e_p$ is the elementary symmetric polynomial of degree $p$; $G = B^T B$ is the Gram matrix of $B$ with respect to the \emph{real} (not the Hermitian) inner product; and $G(\alpha|\alpha)$ is the $(n - r)\times (n - r)$ submatrix of $G$ corresponding to rows and columns complementary to $\alpha$.  These $n$ conditions in fact reduce to $\min(m, n)$ conditions.

In the case of a real symmetric ($CP$-invariant) neutrino mass matrix $M$, the vanishing conditions on the eigenvalues are equivalent to vanishing conditions on the singular values.  Therefore, solving Eq.\ \eqref{conditions} for all $d$ is tantamount to solving our original problem.  While these conditions are complicated to solve in full generality, they simplify in many special cases of interest.

In principle, one could apply the same strategy to find seed textures for complex mass matrices: to determine when $M$ has $m$ vanishing singular values, solve the characteristic equation for $M^\dag M$.  While $x\in \mathbb{C}$ is an eigenvalue of $M$ if and only if $\det(xI - M) = 0$, $x\geq 0$ is a singular value of $M$ if and only if $\det(x^2 I - M^\dag M) = 0$.  However, there exists a simpler strategy.  Solving the eigenvalue conditions \eqref{conditions} yields the most general parametrized textures for real symmetric mass matrices with $m$ vanishing eigenvalues, or rank $n$.  Promoting these parameters to complex numbers results in textures that have rank $n$ as complex matrices, or $m$ vanishing \emph{singular} values.  Hence the real solutions to Eq.\ \eqref{conditions} contain all the information necessary to derive seed textures for \emph{complex} symmetric mass matrices.

\tocless{\section{Mass Matrices}}

To illustrate our formalism, we survey several infinite families of textures that generalize the ordinary and inverse seesaw mechanisms.  These examples represent only a few possible generalizations among many.  Any such texture, after perturbation, gives rise to an explicit pattern of masses and mixings for neutrino oscillations.

We first present some textures that manifestly satisfy the rank requirements, allowing us to bypass solving the algebraic conditions \eqref{conditions}.  Any $N\times N$ matrix of the form $\left[\begin{smallmatrix} 0 & 0 \\ 0 & D \end{smallmatrix}\right]$, where the block $D$ is $n\times n$, clearly has rank at most $n$ and therefore at least $m$ vanishing singular values.  In our language, the ordinary seesaw mechanism can be understood as a perturbation of the off-diagonal blocks:
\begin{equation}
\left[\begin{array}{cc} 0 & 0 \\ 0 & D \end{array}\right]\to \left[\begin{array}{cc} 0 & \times \\ \times & D \end{array}\right],
\label{seesaw}
\end{equation}
which results in $m$ small singular values.

On the other hand, any $N\times N$ matrix of the form $\left[\begin{smallmatrix} 0 & \ast \\ \ast & \ast \end{smallmatrix}\right]$, where the upper $(N - \ell)\times (N - \ell)$ block vanishes, has nullity at least $N - 2\ell$.  If we choose $n$ even and $\ell = n/2$, then such a matrix has at least $m$ vanishing singular values.  Therefore, perturbing such a matrix in a way that preserves the vanishing of the upper $m\times m$ block will result in $m$ small singular values:
\begin{equation}
\left[\begin{array}{ccc} 0 & 0 & \multicolumn{1}{|c}{\ast} \\ 0 & 0 & \multicolumn{1}{|c}{\ast} \\ \cline{1-2} \ast & \ast & \ast \end{array}\right]\to \left[\begin{array}{ccc} 0 & \multicolumn{1}{|c}{\times} & \ast \\ \cline{1-1} \times & \times & \ast \\ \ast & \ast & \ast \end{array}\right],
\label{Lshape}
\end{equation}
where the matrix on the left has a vanishing $(m + n/2)\times (m + n/2)$ block and its perturbation on the right has a vanishing $m\times m$ block.  The entries $\times$ are assumed small relative to $\ast$.  Up to a change of basis within the space of sterile neutrinos, the mechanism \eqref{Lshape} subsumes the inverse seesaw mechanism and its variations \cite{Mohapatra:1986aw, Mohapatra:1986bd, Foot:1988aq, Barr:2003nn, Malinsky:2005bi}.

To construct more explicit textures that fulfill the rank conditions and thus yield realistic neutrino mass matrices upon perturbation, it is convenient to fix the value of $n$.  We require $\operatorname{rank} B < n$ to ensure that $\operatorname{rank} M = n$.  Hence the case of a single sterile neutrino ($n = 1$) requires $B = 0$, reproducing the ordinary seesaw of Eq.\ \eqref{seesaw}.  The simplest case beyond the seesaw is that of two sterile neutrinos ($n = 2$) with $\operatorname{rank} B = 1$ and $\operatorname{rank} M = 2$.  The most general such texture, up to a change of basis that swaps the last two rows and columns, is
\begin{equation}
M = \left[\begin{array}{c|cc}
\mathbf{0}_{m\times m} & \vec{b} & \alpha\vec{b} \\ \hline
\vec{b}^T & \lambda & \mu \\
\alpha\vec{b}^T & \mu & 2\alpha\mu - \alpha^2\lambda
\end{array}\right]
\label{generaln2}
\end{equation}
where all parameters are complex and the $m$-component vector $\smash{\vec{b}}$ is not identically zero.

For arbitrary $n$, rather than classifying all solutions, we focus on the simplest solutions to $\operatorname{rank} M = n$ beyond the seesaw scenario ($B = 0$): those for which $\operatorname{rank} B = 1$.  The most general complex symmetric $M$ of the form \eqref{massmatrix} with $\operatorname{rank} M = n$ and $\operatorname{rank} B = 1$ can be written as
\begin{equation}
M = \left[\begin{array}{cc} 0 & uv^T \\ vu^T & D \end{array}\right],
\label{rank1-1}
\end{equation}
where $u$ and $v$ are nonzero column vectors of length $m$ and $n$, respectively, and
\begin{equation}
\sum_{i=1}^n v_i^2\det(D(i|i)) + 2\sum_{i < j} (-1)^{i + j}v_i v_j\det(D(i|j)) = 0.
\label{rank1-2}
\end{equation}
All parameters are complex.  For example, suppose that only column $i$ of $B$ is nonvanishing:
\begin{equation}
B = \left[\begin{array}{c|c|c|c|c|c|c} \vec{0} & \cdots & \vec{0} & \vec{b} & \vec{0} & \cdots & \vec{0} \end{array}\right], \qquad \vec{b}\neq 0.
\label{onlyonecolumn}
\end{equation}
Then the only requirement that needs to be imposed for $M$ to have rank $n$ is the vanishing of the determinant of the $(n - 1)\times (n - 1)$ submatrix of $D$ obtained by deleting the $i^\text{th}$ row and $i^\text{th}$ column. (The scheme of Eq.\ \eqref{onlyonecolumn} includes the ``cancellation structure'' of Ref.\ \cite{Kersten:2007vk} as a special case.) To aid parameter scans, the solution \eqref{rank1-1}--\eqref{rank1-2} can be parametrized more simply as
\begin{equation}
M = \left[\begin{array}{cc} 0 & u(Uv)^T \\ (Uv)u^T & UDU^T \end{array}\right], \quad \sum_{i=1}^n v_i^2\prod_{j\neq i} d_j = 0,
\end{equation}
where $U$ is a unitary matrix and $D = \operatorname{diag}(d_1, \ldots, d_n)$.

\tocless{\section{Mixing Parameters}}

Experimental data place stringent constraints on the oscillations between light and heavy neutrino states \cite{deGouvea:2015euy, Bolton:2019pcu}.  Consequently, constructing realistic models entails generating not only small active neutrino masses, but also small mixing parameters between active and sterile neutrinos.  While our approach guarantees small masses, the demand for small mixing parameters imposes additional constraints on the resulting mass matrices that can be determined on a case-by-case basis.

To illustrate, we consider the specific example of the texture in Eq.\ \eqref{generaln2} with $\alpha = \lambda = 0$ and all other parameters real.  We assume that the desired perturbation to $M$ will not change the mixing matrix drastically, which allows us to derive an approximation to the mixing matrix analytically. (This is a nontrivial assumption because the eigenvectors of a matrix are generally not continuous functions of its entries, so the mixing matrix of the perturbed mass matrix is not necessarily a small perturbation of that of the unperturbed mass matrix \cite{Kato}.) The orthogonal matrix $O$ for which $O^T MO$ is diagonal takes the block form
\begin{equation}
O = \left[\begin{array}{cc} O_{\nu\nu} & O_{\nu N} \\ O_{N\nu} & O_{NN} \end{array}\right], \quad O_{\nu N} = \frac{\left[\begin{array}{c|c} \vec{b} & -\vec{b} \end{array}\right]}{\sqrt{2(\vec{b}^2 + \mu^2)}}.
\end{equation}
For $\mu\gg |\vec{b}|$, the entries of $O_{\nu N}$ and $O_{N\nu}$ are of order $\vec{b}/\mu$.  Up to phases, these are the off-diagonal mixing parameters.  Therefore, by taking $\mu$ sufficiently large relative to $\smash{\vec{b}}$, we can make all of these components parametrically small.

Note that the hierarchy required by this mechanism is distinct from that of the seesaw.  In the seesaw mechanism, the smallness of masses and mixings is correlated.  Here, by contrast, the smallness of the masses is guaranteed by the structure of the texture itself, and bears no direct relation to the size of the mixing parameters.

\tocless{\section{Phenomenological Applications}}

The aforementioned textures have numerous implications for phenomenology. For instance, they enable significant mixing between active and sterile neutrinos beyond that allowed by the seesaw mechanism as a result of effectively decoupling the smallness of the masses and the mixing parameters. Correspondingly, they offer potential solutions to the anomalies observed in the LSND \cite{LSND:1996ubh, LSND:2001aii} and MiniBooNE \cite{MiniBooNE:2008hfu} experiments. (For an application of a special case of the texture in Eq.\ \eqref{generaln2}, see Ref.\ \cite{Babu:2022non}.)

Another intriguing prospect is that such textures may address the $(g-2)_\mu$ discrepancy, substantially impacting lepton flavor violation \cite{Gonzalez-Garcia:1988okv, Coy:2018bxr} and collider phenomenology \cite{Pilaftsis:1991ug, Kersten:2007vk}. In the following discussion, we focus on the left-right symmetric model (LRSM) \cite{Pati:1974yy, Mohapatra:1974gc, Mohapatra:1974hk, Senjanovic:1975rk, Senjanovic:1978ev, Mohapatra:1979ia, Mohapatra:1980yp, Wyler:1982dd}, based on the gauge group $SU(2)_L \times SU(2)_R \times U(1)_{B-L}$, as a potential solution to this $(g-2)_\mu$ anomaly. A crucial difference between our analysis and that of previous work \cite{Boyarkina:2003lez, Boyarkin:2008zz, Taibi:2015ura} is that the explicit realization of our textures allows us to move beyond standard seesaw assumptions.

The mismatch between experimental and theoretical determinations of the anomalous magnetic moment of leptons $a_\ell = (g-2)_\ell/2$, which in the SM is calculated perturbatively in the fine-structure constant $\alpha_\text{em}$, hints at physics beyond the SM. The Muon $g-2$ Collaboration at Fermilab \cite{Abi:2021gix, Muong-2:2023cdq} has confirmed the long-standing \cite{Bennett:2006fi} discrepancy $\Delta a_\mu = a_\mu(\text{experiment}) - a_\mu(\text{theory}) = (2.49\pm 0.48)\times 10^{-9}$ at a combined $5.1\sigma$ deviation from the SM prediction (see Ref.\ \cite{Aoyama:2020ynm} and references therein). Lattice QCD results \cite{Borsanyi:2020mff} are in tension with the SM theory prediction. With no consensus yet in the community, we take the deviation at face value and examine its potential resolution within the LRSM.

\textbf{Model.} The minimal left-right symmetric model naturally incorporates the right-handed neutrino $\nu_R$ within a right-handed lepton doublet, enabling neutrino mass generation through the seesaw mechanism \cite{Mohapatra:1979ia, Mohapatra:1980yp}. It explains parity violation through the phenomenon of spontaneous gauge symmetry breaking, with the promotion of hypercharge $Y$ to $B-L$ offering insights into its origin from higher unification (e.g., $SO(10)$). Moreover, if realized around the TeV scale, this model can be tested in ongoing and future low-energy and high-energy collider experiments. The Higgs sector comprises the fields $\Delta_L (3, 1, 2) + \Delta_R(1, 3, 2) + \Phi(2, 2, 0)$:
\begin{equation}
\Phi = \begin{pmatrix}
\phi_1^0 & \phi_1^+ \\
\phi_2^- & \phi_2^0
\end{pmatrix}, \quad
\Delta_{L,R} = \begin{pmatrix}
\frac{\delta^+}{\sqrt{2}} & \delta^{++} \\
\delta^0 & -\frac{\delta^+}{\sqrt{2}}
\end{pmatrix}_{L, R}.
\end{equation}
After the neutral component of $\Delta_R$ develops a VEV $\langle\delta_R^0\rangle \linebreak[1] = \linebreak[1] \smash{v_R/\sqrt{2}}$, the $SU(2)_R$ symmetry is broken, giving masses to the $W_R^\pm$ and $Z_R$ gauge bosons. The VEVs $\langle\phi_{1,2}^0\rangle = \smash{\kappa_{1,2}/\sqrt{2}}$ break the remaining $SU(2)_L\times U(1)_{B-L}$ down to the usual $U(1)_\text{em}$, thereby setting the mass scale for $SU(2)_L$ gauge bosons, with $\kappa_1^2 + \kappa_2^2\simeq 246^2$ GeV$^2$. Note that the VEVs should obey $v_R \gg \kappa_1, \kappa_2$ to satisfy con\-straints from low-energy weak interactions. The most general Yukawa Lagrangian of the model is given by
\begin{align}
\mathcal{L}_Y &= \overline{Q}_L (Y \Phi + \widetilde{Y} \widetilde{\Phi})  Q_R + \overline{\psi}_L (y \Phi + \widetilde{y} \widetilde{\Phi}) \psi_R \nonumber \\
&+ f(\psi_L^T C i\sigma_2 \Delta_L \psi_L + \psi_R^T C i\sigma_2 \Delta_R \psi_R) + \text{h.c.}, \label{eq:LagLR}
\end{align}
where $\psi_{L, R}$ ($Q_{L, R}$) are the lepton (quark) doublets, $\smash{\widetilde{\Phi}} = \linebreak[1] \sigma_2 \Phi^\ast \sigma_2$, and $\sigma_2$ is the second Pauli matrix. Under left-right parity symmetry, the fermion and scalar fields transform as $\Phi \leftrightarrow \Phi^\dagger$, $\smash{\widetilde{\Phi}} \leftrightarrow \smash{\widetilde{\Phi}}^\dagger$, $\Delta_L \leftrightarrow \Delta_R$, $\Psi_L \leftrightarrow \Psi_R$, and $Q_L \leftrightarrow Q_R$. The VEVs for the Higgs fields generate fer\-mion masses. The mass matrices for charged leptons ($M_\ell$) and Dirac neutrinos ($M_{\nu_D}$) are given by
\begin{alignat}{2}
% M_u &= \frac{Y \kappa_1 + \widetilde{Y} \kappa_2}{\sqrt{2}}, \quad & M_d &= \frac{Y \kappa_2 + \widetilde{Y} \kappa_1}{\sqrt{2}}, \label{eq:dM} \\[5pt]
M_\ell &= \frac{y \kappa_2 + \widetilde{y} \kappa_1}{\sqrt{2}}, \quad & M_{\nu_D} &= \frac{y \kappa_1 + \widetilde{y} \kappa_2}{\sqrt{2}}, \label{eq:nuM}
\end{alignat}
while those for up and down quarks are given by $M_{u, d} = (Y \kappa_{1, 2} + \smash{\widetilde{Y}} \kappa_{2, 1})/\sqrt{2}$. The $6\times 6$ mass matrix for the $\nu$-$N$ sector (with $\nu \equiv \nu_L$ and $N\equiv C (\overline{\nu}_R)^T$) is
\begin{equation}
M_\nu = \begin{pmatrix}
0 & M_{\nu_D} \\
M_{\nu_D}^T & M_{\nu_R}
\end{pmatrix},
\label{eq:numat}
\end{equation}
where $M_{\nu_R} = \sqrt{2} f v_R$ and the upper-left block of Eq.\ \eqref{eq:numat} follows from $\langle \delta_L^0 \rangle = 0$. The relations in Eq.\ \eqref{eq:nuM} can be inverted, in the limit $\kappa_1 \neq \kappa_2$, to express the Yukawa coupling matrices in terms of the mass matrices:
\begin{equation}
y = \frac{\sqrt{2}(\kappa_1 M_{\nu_D} - \kappa_2 M_\ell)}{\kappa_1^2 - \kappa_2^2}, \quad \widetilde{y} = \frac{\sqrt{2}(\kappa_1 M_{\ell} - \kappa_2 M_{\nu_D})}{\kappa_1^2 - \kappa_2^2}. \label{eq:yuk1}
\end{equation}
A similar inversion can be done for the Yukawa couplings $Y$ and $\smash{\widetilde{Y}}$. Without loss of generality, we take $|\kappa_2/\kappa_1| \leq 1$, where the ratio is constrained by perturbativity. For instance, if one requires the top quark Yukawa coupling to be $\leq 1.5$, then the upper limits $(0.578, 0.616, 0.645)$ on $|\kappa_2/\kappa_1|$ correspond to the LR breaking scale $v_R$ being $(1, \linebreak[1] 10, \linebreak[1] 100)$ TeV \cite{Babu:2020bgz}. These values are derived by evolving the top quark Yukawa coupling to the scale $v_R$.

Among the Higgs bosons, $\phi_{1,2}^+$ are of primary interest here because they provide chirally enhanced contributions to $(g-2)_\mu$. The transformation of the charged scalars $\{\phi_{1,2}^+, \linebreak[1] \delta_{L, R}^+\}$ from the flavor basis to the mass basis $\{G_{L, R}^+, \linebreak[1] \smash{h_{1, 2}^+}\}$ can be written as
\begin{equation}
\phi_{1, 2}^+ \simeq \frac{\kappa_{1, 2} h_2^+ \mp \kappa_{2, 1} G_L^+}{\sqrt{\kappa_1^2 + \kappa_2^2}}, \quad \delta_R^+ \simeq G_R^+, \quad \delta_L^+ \simeq h_1^+,
\end{equation}
where $G_{L, R}^\pm$ are the massless Goldstone modes associated with the massive gauge bosons $\smash{W_{L, R}^\pm}$. A detailed analysis of the scalar sector can be found in Ref.\ \cite{Deshpande:1990ip}.

\begin{figure}[!htb]
\centering
\includegraphics[scale=0.4]{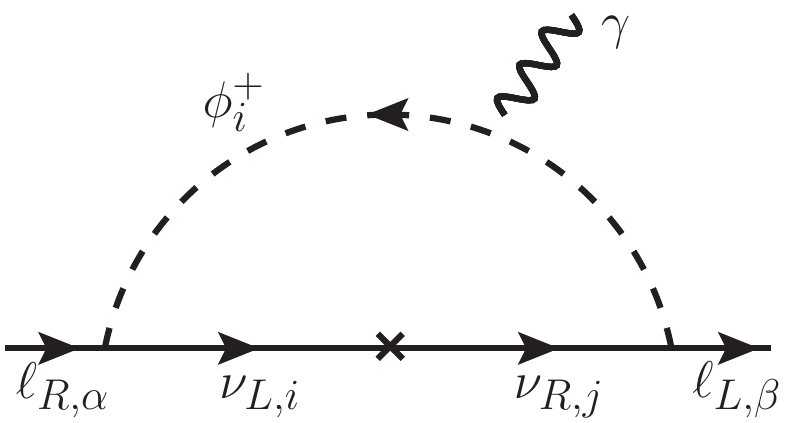}
\caption{One-loop Feynman diagram contributing to anomalous magnetic moment ($\alpha = \beta$) and lepton-flavor-violating decays ($\alpha \neq \beta$) through the heavy charged scalar $\phi_i$.}
\label{fig:oneloop}
\end{figure}

\begin{figure}[!htb]
\centering
\vspace{2mm}
\includegraphics[width=0.48\textwidth]{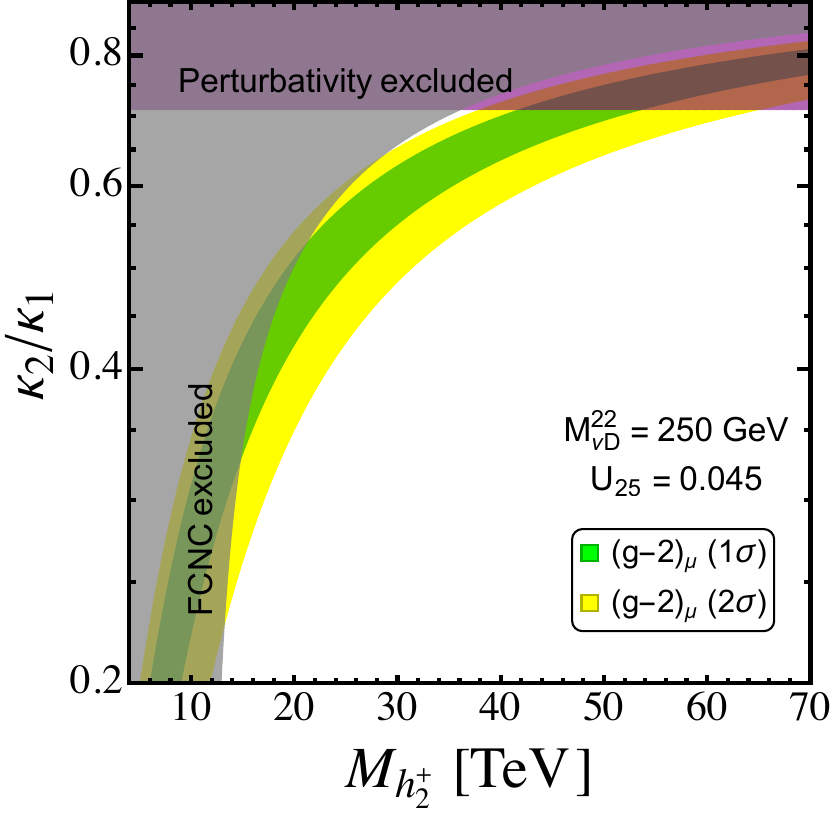}
\caption{The parameter space of $\kappa_2/\kappa_1$ versus scalar mass $\smash{M_{h_2^+}}$. The green and yellow bands indicate the regions allowed by the $(g-2)_\mu$ constraints to $1\sigma$ and $2\sigma$, respectively. The region in gray is excluded by FCNC constraints \cite{Kiers:2002cz, Zhang:2007da}, whereas the region in purple is excluded by perturbativity.}
\label{fig:g2plot}
\end{figure}

\textbf{Anomalous magnetic moment.} Quantum corrections due to charged scalars and the neutrino chirality flip can modify the electromagnetic interactions of charged leptons, as depicted in Fig.\ \ref{fig:oneloop}. Diagrams without the chirality flip as well as those from the gauge sector are subdominant and not considered here \cite{Boyarkin:2008zz, Taibi:2015ura}. We choose the following neutrino mass matrix texture for Eq.\ \eqref{eq:numat} to reproduce the light neutrino masses and mixings and to get a chirally enhanced contribution to $(g-2)_\mu$:
\begin{equation}
M_\nu = \left[\begin{array}{ccc|ccc}
0 & 0 & 0 & \times & \times & \times \\
0 & 0 & 0 & \times & M_{\nu_D}^{22} & \times \\
0 & 0 & 0 & \times & \times & \times \\ \hline
\times & \times & \times & m_{11} & m_{12} & m_{13} \\
\times & M_{\nu_D}^{22} & \times & m_{12} & m_{22} & m_{23} \\
\times & \times & \times & m_{13} & m_{23} & m_{33}
\end{array}\right],
\label{exampletexture}
\end{equation}
where the entries marked by $\times$ are much smaller than $M_{\nu_D}^{22}$. Choosing $M_{\nu_D}^{22}$ of order 100 GeV has a negligible impact on the light neutrino masses, provided that $m_{11} m_{33} - m_{13}^2 = 0$ (as per the discussion below Eq.\ \eqref{onlyonecolumn}). Note that this differs from the usual seesaw setup. Denoting the heavy neutrino masses by $M_{N_i}$ and the mixing parameters between the light $\nu_\mu$ and the sterile states $N_i$ by $U_{2i}$ where $i\in \{4, 5, 6\}$, to one-loop order, the contribution of species $i$ to $(g-2)_\mu$ takes the form \cite{Leveille:1977rc}
\begin{align}
\Delta a_\mu &\simeq \frac{2 m_\mu}{16 \pi^2} \frac{M_{N_i} U_{2i}}{M_{h_2^+}^2} \frac{(y\kappa_1 - \widetilde{y}\kappa_2) (y\kappa_2 - \widetilde{y}\kappa_1)}{\kappa_1^2 + \kappa_2^2}\, F\left[\frac{M_{N_i}^2}{M_{h_2^+}^2}\right] \notag \\
&\text{ where } F[x]\equiv \frac{1}{1 - x^2}(1 - x^2 + 2x\log x). \label{onelooporder}
\end{align}
Here, $y$ and $\widetilde{y}$ can be substituted in favor of $M_{\nu D}^{22}$ using Eq.\ \eqref{eq:yuk1}. The sign of the contribution \eqref{onelooporder} can be arbitrary due to the different Yukawa couplings appearing in the loop diagram, unlike for the charged scalar contribution without the chirality flip. Rewriting the above equation in terms of $\{\smash{M_{h_2^+}}, U_{2 i}, M_{\nu D}^{22}, \kappa_2/\kappa_1\}$, the allowed parameter space of the model is shown in Fig.\ \ref{fig:g2plot}. Note that in Fig.\ \ref{fig:g2plot}, we choose the specific value $i = 5$, and we also assume that the chosen texture respects the relation $U_{25} \approx M_{\nu_D}^{22}/M_{N_5}$, which is similar to (but more specific than) what would arise in the seesaw approximation. Notably, unlike in the seesaw framework where $U_{2 i}$ must be small to achieve tiny neutrino masses, it can take arbitrarily large values here. We choose the mixing to be $0.045$ to satisfy the active-sterile mixing bound \cite{deGouvea:2015euy, Bolton:2019pcu}.

\tocless{\section{Discussion}}

We have demonstrated a systematic method to construct neutrino mass matrices through the perturbation of certain ``seed'' textures, which encompasses all neutrino mass mechanisms involving sterile neutrinos. The proposed approach does not require a hierarchy between Dirac and Majorana masses, thus offering new avenues toward models of neutrino mass generation that could be probed in experiments.

One challenge for future work is to understand or categorize the symmetries of the textures that arise in our framework from a top-down perspective. By motivating the smallness of the perturbations to such textures via symmetry breaking, one may lessen or obviate the need for fine-tuning \cite{Kersten:2007vk}. Another challenge for future work is to implement this method in a way that efficiently accounts for constraints on active-sterile neutrino mixing. Some additional technical issues that deserve further study include placing precise bounds on the mass eigenvalues obtained by perturbation theory \cite{Kato} and exploring redundancies in parametrizations of the mass matrix (along the lines of Ref.\ \cite{Heeck:2012fw}, which works solely within the framework of the type I seesaw).

Finally, we note that the philosophy of our work extends far beyond the specific ansatz \eqref{massmatrix} or even the assumption of an active-sterile neutrino mass hierarchy. For example, given a texture of the form \eqref{massmatrix}, small bare Majorana masses for left-handed neutrinos can be incorporated by applying first-order perturbation theory to a perturbation of the form $\left[\begin{smallmatrix} A & 0 \\ 0 & 0 \end{smallmatrix}\right]$. Bare masses of any size can be accommodated by applying our rank conditions to a completely general ansatz for $M$. Furthermore, it is often useful to consider pseudo-Dirac scenarios with $B\gg D$ in which active and sterile neutrinos are nearly degenerate in mass \cite{Wolfenstein:1981kw, Petcov:1982ya, Valle:1983dk}. Our approach suggests a broad generalization of such scenarios in which nearly degenerate masses arise from perturbing a seed texture with an \emph{exactly} degenerate spectrum of singular values (rather than with a certain rank, as considered here). All of these scenarios open new possibilities for investigation.

\tocless{\section{Acknowledgements}}

YF thanks C.\ Kilic for a useful conversation. We acknowledge the Aspen Center for Physics for hospitality during the initial stages of this project. Work at the Aspen Center was supported by the National Science Foundation under Grant No.\ PHY-1607611. The work of YF was supported in part by the NSF grants PHY-1914679 and PHY-2210562. The work of AT was supported in part by the NSF grant PHY-2210428.

This article has been authored by an employee of National Technology \& Engineering Solutions of Sandia, LLC under Contract No.\ DE-NA0003525 with the U.S. Department of Energy (DOE). The employee owns all right, title, and interest in and to the article and is solely responsible for its contents. The U.S. Government retains, and the publisher, by accepting the article for publication, acknowledges that the U.S. Government retains, a non-exclusive, paid-up, irrevocable, worldwide license to publish or reproduce the published form of this article or allow others to do so, for U.S. Government purposes. The DOE will provide public access to these results of federally sponsored research in accordance with the DOE Public Access Plan \href{https://www.energy.gov/downloads/doe-public-access-plan}{https://www.energy.gov/downloads/doe-public-access-plan}.

% \nocite{*}
\bibliographystyle{utphys}
\tocless{\bibliography{references}}

\onecolumngrid
\appendix

\newpage

\tableofcontents

\section{Background}

\subsection{Seesaw Mechanisms}

The accidental lepton number symmetry (or the exact $B - L$ symmetry) of the Standard Model forbids Majorana masses for left-handed neutrinos.  However, such masses can emerge through various mechanisms in extensions of the SM.  In effective field theory, Weinberg's dimension-five operator \cite{Weinberg:1979sa} generates Majorana masses for left-handed neutrinos after electroweak symmetry breaking; the seesaw mechanism is a particular realization of this operator at tree level.  Alternatively, left-handed Majorana masses can be generated at the renormalizable level by Higgs triplets rather than doublets.

The standard realization of the seesaw mechanism introduces sterile (gauge-neutral) neutrinos\footnote{In this paper, we use the terms ``active'' and ``sterile'' to refer to light and heavy mass eigenstates, respectively, rather than the flavor eigenstates that are charged or neutral under the SM gauge group.  In the seesaw approximation, where the mass eigenstates involve very little mixing between left- and right-handed flavor eigenstates, these two notions are nearly synonymous, but this is not necessarily so in our more general setting.} to the SM with Yukawa couplings to the left-handed lepton doublets and the SM Higgs doublet, giving Dirac masses to the neutrinos in the same way as the SM fermions.  Since these sterile neutrinos are gauge singlets, a Majorana mass term that violates lepton number is also allowed.  Note that such mass terms can be forbidden by assuming additional gauge or global symmetries, making the neutrino a Dirac fermion.  Introducing three right-handed neutrinos without any additional symmetry gives rise to Weinberg's dimension-five operator, $LLHH/\Lambda$, where $\Lambda$ is the scale of the right-handed neutrino.  The Yukawa coupling becomes a Dirac mass term after electroweak symmetry breaking.  Denoting the Dirac mass by $m_{LR}$ and the Majorana mass by $M_{RR}$, in the limit that $m_{LR}\ll M_{RR}$, the masses of the light and heavy states go like $-m_{LR}^2/M_{RR}$ (observed) and $M_{RR}$ (unobserved).  This is popularly known as the type I seesaw mechanism.

We catalog below some specific realizations of the seesaw mechanism.  We denote by $\psi_{L, R}$ the left- and right-handed components of the Dirac spinor $\psi$, and by $\psi_{L, R}^c\equiv (\psi_{L, R})^c$ their $CP$-conjugates.  In all of the following scenarios, the light neutrino masses can be estimated by integrating out the heavy mass eigenstates.

$\bullet$ The ordinary seesaw mechanism stems from the observation that for a $2\times 2$ matrix $\left[\begin{smallmatrix} a_{11} & a_{12} \\ a_{12} & a_{22} \end{smallmatrix}\right]$ with $a_{11}\ll a_{12}\ll a_{22}$, the eigenvalues are approximately $a_{22}$ and $a_{11} - a_{12}^2/a_{22}$.  In the type I seesaw mechanism, Majorana masses for the left-handed neutrinos are assumed to be absent in the fundamental theory, but are induced by masses for right-handed neutrinos \cite{Minkowski:1977sc, Yanagida:1979as, Gell-Mann:1979vob, Glashow:1979nm, Mohapatra:1979ia, Schechter:1980gr}.  In this case, the $6\times 6$ mass matrix takes the form
\begin{equation}
\left[\begin{array}{cc} 0 & m_{LR} \\ m_{LR}^T & M_{RR} \end{array}\right]
\end{equation}
in the flavor basis $(\nu_L^c, \nu_R)$, which arises from the Lagrangian
\begin{equation}
\mathcal{L} = -\overline{\nu_L}m_{LR}\nu_R - \frac{1}{2}\nu_R^T M_{RR}\nu_R + \text{h.c.}
\end{equation}
In the limit that $M_{RR}\gg m_{LR}$, integrating out $\nu_R$ gives
\begin{equation}
\mathcal{L} = -\frac{1}{2}\overline{\nu_L}m_{LL}\nu_L^c + \text{h.c.}, \qquad m_{LL} = -m_{LR}M_{RR}^{-1}m_{LR}^T.
\end{equation}
In the type II seesaw mechanism, the bare left-handed Majorana masses are allowed to be nonzero \cite{Mohapatra:1979ia, Schechter:1980gr, Magg:1980ut}. (This is sometimes called a ``mixed seesaw'' when the bare $m_{LL}\neq 0$, and type II specifically when the contribution of $m_{LL}$ to the light neutrino masses dominates \cite{Mohapatra:2006gs}.) The type III seesaw mechanism, on the other hand, utilizes $SU(2)_L$ triplet fermions instead of a singlet \cite{Foot:1988aq}.

$\bullet$ The inverse (or double) seesaw mechanism (ISS) \cite{Mohapatra:1986aw, Mohapatra:1986bd, Barr:2003nn} involves the $9\times 9$ mass matrix
\begin{equation}
\left[\begin{array}{ccc} 0 & m_{LR} & 0 \\ m_{LR}^T & 0 & M_{RS} \\ 0 & M_{RS}^T & M_{SS} \end{array}\right]
\label{ISS}
\end{equation}
in the basis $(\nu_L^c, \nu_R, S)$ with $S$ being an additional singlet, which arises from the Lagrangian
\begin{equation}
\mathcal{L} = -\overline{\nu_L}m_{LR}\nu_R - \nu_R^T M_{RS}S - \frac{1}{2}S^T M_{SS}S + \text{h.c.}
\end{equation}
Integrating out $\nu_R$ and $S$ gives
\begin{equation}
\mathcal{L} = -\frac{1}{2}\overline{\nu_L}m_{LL}\nu_L^c + \text{h.c.}, \qquad m_{LL} = m_{LR}M_{RS}^{-T}M_{SS}M_{RS}^{-1}m_{LR}^T.
\end{equation}
This procedure is valid as long as $M_{RS}\gg m_{LR}$, in which case there exist one light mass eigenstate and two heavy mass eigenstates, with the former being close to $\nu_L$ and the latter being combinations of $\nu_R$ and $S$.  The resulting $m_{LL}$ is then approximately the matrix of light neutrino masses.  This model achieves a double suppression of the mass scale, hence the alternative name double seesaw.  The linear seesaw (LSS), by contrast, includes only the off-diagonal block perturbations $M_{LS}$ in \eqref{ISS}, setting $M_{SS} = 0$ \cite{Malinsky:2005bi}. (Note that the type III seesaw allows for nonzero $M_{LS}$ in \eqref{ISS} \cite{Foot:1988aq}.)

$\bullet$ Finally, all of the aforementioned seesaw mechanisms occur as special cases of the texture \cite{Dev:2012sg, Chakraborty:2014hfa, Sinha:2015ooa, Han:2021qum}
\begin{equation}
\left[\begin{array}{ccc} 0 & m_{LR} & M_{LS} \\ m_{LR}^T & M_{RR} & M_{RS} \\ M_{LS}^T & M_{RS}^T & M_{SS} \end{array}\right],
\end{equation}
which arises from the Lagrangian
\begin{equation}
\mathcal{L} = -\overline{\nu_L}m_{LR}\nu_R - \overline{\nu_L}M_{LS}S - \nu_R^T M_{RS}S - \frac{1}{2}\nu_R^T M_{RR}\nu_R - \frac{1}{2}S^T M_{SS}S + \text{h.c.}
\end{equation}
Assuming the existence of a single light mass eigenstate that is close to $\nu_L$, we can integrate out $\nu_R$ and $S$ (or more precisely, the heavy modes) to obtain $\mathcal{L} = -\frac{1}{2}\overline{\nu_L}m_{LL}\nu_L^c + \text{h.c.}$ where
\begin{equation}
m_{LL} = -(m_{LR} - M_{LS}M_{SS}^{-1}M_{RS}^T)(M_{RR} - M_{RS}M_{SS}^{-1}M_{RS}^T)^{-1}(m_{LR}^T - M_{RS}M_{SS}^{-1}M_{LS}^T) - M_{LS}M_{SS}^{-1}M_{LS}^T.
\end{equation}
This is an approximation to the matrix of light neutrino masses.

\subsection{Physical Neutrino Masses}

Given a complex symmetric neutrino mass matrix $M$, the physical neutrino masses correspond to the singular values of $M$, which generally differ from the eigenvalues of $M$.  Indeed, the mass term in the Lagrangian takes the form $\nu_f^T M\nu_f$, where $\nu_f$ is a vector of flavor eigenstates, and basis changes are effected by unitary transformations of $\nu_f$.  The Autonne-Takagi factorization \cite{Autonne, Takagi} states that there exists a unitary matrix $U$ such that $\smash{U^T MU}$ is diagonal with nonnegative real entries:
\begin{equation}
U^T MU = M_D.
\label{ATfactorization}
\end{equation}
The diagonal entries of $M_D$ are the singular values of $M$, or the nonnegative square roots of the eigenvalues of $M^\dag M$.  Indeed, $U^\dag M^\dag MU = M_D^2$.  This ``diagonalization'' in the Autonne-Takagi sense should be contrasted with diagonalization in the traditional sense, i.e.,
\begin{equation}
P^{-1}MP = D
\label{diagonalization}
\end{equation}
where the nontrivial entries of the diagonal matrix $D$ are the eigenvalues of $M$.  Any complex symmetric matrix admits a factorization of the form \eqref{ATfactorization} (which guarantees the existence of a basis of ``mass eigenstates'' with nonnegative real masses), but is not necessarily diagonalizable in the sense \eqref{diagonalization}.  However, if $M$ is \emph{real} and symmetric (hence Hermitian), then it is diagonalizable by a real orthogonal matrix:
\begin{equation}
O^T MO = D.
\end{equation}
In this case, the singular values of $M$ are the absolute values of the (real) eigenvalues of $M$.  A suitable $U$ in \eqref{ATfactorization} can be found by multiplying $O$ in \eqref{diagonalization} on the right by a diagonal matrix of phases if necessary.  If $M$ is additionally positive-semidefinite, then the factorization \eqref{ATfactorization} and the diagonalization \eqref{diagonalization} coincide.

A congruence transformation $\cdot\mapsto U^T\cdot U$ (or more generally, multiplication by any invertible matrix) preserves rank and nullity.  However, it need not preserve the number of zero eigenvalues.  This is because the geometric multiplicity of an eigenvalue is bounded above by its algebraic multiplicity, and in particular, the nullity is only bounded above by the number of zero eigenvalues.  So if $M$ is not diagonalizable, then the number of vanishing singular values may be smaller than the number of vanishing eigenvalues of $M$.  However, the rank of $M^\dag M$ (which is Hermitian and hence diagonalizable) is the rank of $M$, so if $M$ has rank at most $N - m = n$, then it will have at least $m$ vanishing eigenvalues \emph{and} singular values.  The nullity of $M$ is the number of vanishing singular values.

If a matrix is normal (or equivalently, unitarily diagonalizable), then the singular values are simply the absolute values of the eigenvalues.  A vanishing singular value is also an eigenvalue, but the converse does not hold.  In general, there is no direct relation between the singular values and the eigenvalues of a matrix.  However, the singular values do yield bounds on the eigenvalues.  Let the $N\times N$ complex matrix $M$ have singular values $\sigma_i$ ordered as $\sigma_1\geq \cdots\geq \sigma_N\geq 0$ and eigenvalues $\lambda_i$ ordered as $|\lambda_1|\geq \cdots\geq |\lambda_N|$.  The Weyl-Horn inequalities \cite{Weyl, Horn} state that
\begin{equation}
\prod_{i=1}^k |\lambda_i|\leq \prod_{i=1}^k \sigma_i
\end{equation}
for $k = 1, \ldots, N$, with equality for $k = N$.  In particular, the absolute values of the eigenvalues lie between the largest and smallest singular values.  These inequalities generate the complete set of relations between singular values and eigenvalues.  For example, it follows that
\begin{equation}
\sum_{i=1}^k |\lambda_i|^p\leq \sum_{i=1}^k \sigma_i^p
\end{equation}
for $k = 1, \ldots, N$ and any $p\geq 0$, i.e., the ($p^\text{th}$ powers of the) singular values weakly majorize the ($p^\text{th}$ powers of the) absolute values of the eigenvalues.  This is a special case of a stronger result, Weyl's Majorant Theorem \cite{Weyl}.  See \cite{Bhatia} for a more comprehensive overview of the theory.

\subsection{Mixing Parameters}

With $n$ sterile neutrinos, the flavor basis and the mass basis are related by a $(3 + n)\times (3 + n)$ unitary matrix $U$.  The upper $3\times 3$ block of this full mixing matrix is the PMNS matrix, which is tightly constrained by experiment.  The off-diagonal blocks, and in particular the mixing parameters $U_{1N}, U_{2N}, U_{3N}$ where $N$ is any species of sterile neutrino, are likewise constrained by experiment.  In the seesaw mechanism, the magnitude of the block $U_{\nu N}$ goes like $B/D$, which can be tuned to satisfy these constraints.  More precisely, given a mass matrix of the form \eqref{massmatrix}, the mixing matrix takes the following block form \cite{Schechter:1981cv, Coy:2018bxr, Flieger:2020lbg}:
\begin{equation}
U = \left[\begin{array}{cc} 1 - \frac{1}{2}B^\dag(DD^\dag)^{-1}B & (D^{-1}B)^\dag \\ -D^{-1}B & 1 - \frac{1}{2}D^{-1}BB^\dag(D^\dag)^{-1} \end{array}\right] + O((D^{-1}B)^3).
\label{seesawmixing}
\end{equation}
In this way, the seesaw mechanism correlates the sizes of the masses and the mixing parameters.

Restricting our attention to the light neutrinos, and assuming unitarity of the PMNS matrix $U_\text{PMNS}$, the flavor basis $\nu_f$ ($\nu_{\alpha = e, \mu, \tau}$) and the mass basis $\nu$ ($\nu_{i = 1, 2, 3}$) are related by $\nu_f = U_\text{PMNS}\nu$.  Hence the mass matrix $m_\nu$ of the light neutrinos is diagonalized as follows:
\begin{equation}
m_\text{diag} = U_\text{PMNS}^T m_\nu U_\text{PMNS}.
\end{equation}
Taking into account the symmetries of the lepton sector, $U_\text{PMNS}$ can be parametrized by three mixing angles and one $CP$-violating phase $\delta_{CP}$.  Moreover, assuming that the $\nu$ are Majorana fermions, there exist relative phases among the Majorana masses $m_1, m_2, m_3$.  Choosing $m_3$ to be real and positive, we may absorb these phases into $m_{1, 2}\equiv |m_{1, 2}|e^{\phi_{1, 2}}$.  Thus the inclusion of nonzero neutrino masses adds nine parameters to the Standard Model: three physical masses, three mixing angles, and three $CP$-violating phases \cite{Altarelli:2004za}.  The Majorana $CP$-violating phases are not experimentally accessible, which is why one typically focuses on the Dirac $CP$-violating phase $\delta_{CP}$.

Letting $\Delta m_{ij}^2\equiv m_i^2 - m_j^2$ (and replacing $m_i$ with $|m_i|$ as necessary for Majorana masses), neutrino oscillation data are compatible with a normal hierarchy $m_3\gg m_2 > m_1$ (with $\smash{\Delta m_{32}^2} > 0$), an inverted hierarchy $m_2 > m_1\gg m_3$ (with $\Delta m_{32}^2 < 0$), or an approximate degeneracy ($m_1\approx m_2\approx m_3$) \cite{King:2003jb, Altarelli:2004za, Mohapatra:2006gs}.

\section{Examples}

In this appendix, we elaborate on some explicit examples presented in the main text.

\subsection{Beyond the Seesaw}

We first describe how our rank conditions for $m$ vanishing singular values yield neutrino mass matrix textures that generalize those of the ordinary and inverse seesaw mechanisms.

The ordinary seesaw mechanism relies on perturbing the off-diagonal blocks of the texture
\begin{equation}
\left[\begin{array}{cc} 0 & 0 \\ 0 & D \end{array}\right],
\label{seesawtexture}
\end{equation}
where $D$ is $n\times n$, while keeping the upper-left $m\times m$ block zero.  This texture clearly has rank at most $n$ and therefore at least $m$ vanishing singular values.

The inverse seesaw mechanism involves a perturbation of a $9\times 9$ mass matrix with $3\times 3$ blocks,
\begin{equation}
\left[\begin{array}{ccc} 0 & \mu & 0 \\ \mu^T & M'' & M' \\ 0 & M'^T & 0 \end{array}\right],
\label{crosstexture}
\end{equation}
where $M''$ is symmetric (in fact, the block $M''$ is typically set to zero).  Any such texture has a nullspace of dimension at least three and therefore at least three vanishing singular values.  To obtain three small masses, we can perturb the lower-right 0 block (as in the inverse seesaw) as well as the off-diagonal 0 blocks (as in the type III seesaw).

More generally, let $m, n$ be arbitrary and choose $\ell\leq n$.  Then consider the ``cross'' texture \eqref{crosstexture} where now:
\[
\text{$\mu$ is an $m\times \ell$ block}, \qquad \text{$M'$ is an $\ell\times (n - \ell)$ block}, \qquad \text{$M''$ is an $\ell\times \ell$ block}.
\]
Such a texture has nullity at least $m + n - 2\ell$.  So choosing $n$ even and $\ell = n/2$ yields a texture with at least $m$ vanishing singular values.  Perturbing such a texture will give rise to $m$ light neutrinos.  The inverse seesaw mechanism corresponds to the special case $(m, n) = (3, 6)$, $\ell = 3$, and $M'' = 0$.

To state the above in a more illuminating way, by a change of basis (conjugation by elementary matrices), the cross texture \eqref{crosstexture} is equivalent to a texture of the form
\begin{equation}
\left[\begin{array}{cc} 0 & \mu \\ \mu^T & M' \end{array}\right]
\label{simplecross}
\end{equation}
where $\mu$ is $(m + n - \ell)\times \ell$ and $M'$ is $\ell\times \ell$ (the $\mu$ and $M'$ in \eqref{crosstexture} and \eqref{simplecross} are different).  It is easy to see that such a matrix has a nullspace of dimension at least $m + n - 2\ell$.\footnote{For instance, any $(m + n)\times (m + n)$ block matrix $A = \left[\begin{smallmatrix} 0 & B \\ C & D \end{smallmatrix}\right]$ satisfies
\[
A\left[\begin{smallmatrix} \vec{v} \\ \vec{w} \end{smallmatrix}\right] = \left[\begin{smallmatrix} B\vec{w} \\ C\vec{v} + D\vec{w} \end{smallmatrix}\right],
\]
so the first $m$ entries of any vector in the image of $A$ are constrained to lie in the span of the $n$ columns of $B$.  Hence the nullspace of $A$ is at least $(m - n)$-dimensional.}  Choose $n$ even and $\ell = n/2$.  Then the inverse seesaw mechanism relies on perturbing the above texture so that $\mu$ becomes $m\times 2\ell$ and $M'$ becomes $2\ell\times 2\ell$.  In fact, the inverse seesaw mechanism would have the original $M' = 0$; our generalization allows for arbitrary $M'$.  Moreover, one can make a change of basis in the space of sterile neutrinos after the perturbation so that the structure or even the presence of perturbations is not manifest.

Our prescription for deriving neutrino mass matrices by perturbing matrices that satisfy certain rank conditions therefore generalizes both the ordinary and inverse seesaw mechanisms.  In fact, on general grounds, it generalizes \emph{any} seesaw mechanism because it provides necessary and sufficient conditions for the generation of small active neutrino masses regardless of whether there exists a hierarchy between Dirac and Majorana masses for sterile neutrinos.

\subsection{Rank Conditions}

We now demonstrate how to solve the rank conditions to derive explicit textures (either real or complex) that give rise to realistic neutrino mass matrices upon perturbation.

Note that for a single sterile neutrino ($n = 1$), this prescription yields nothing beyond the ordinary seesaw mechanism.  Indeed, consider a mass matrix $M$ of the form \eqref{massmatrix} with $n = 1$.  If $B$ is nonzero, then $M$ has rank two regardless of the value of $D = [d]$.  Hence the only way to arrange for $M$ to have rank one is to set $B = 0$ and $d\neq 0$, corresponding (upon perturbation) to the ordinary seesaw mechanism.  Of course, the case $n = 1$ is not phenomenologically viable because it can only produce one light neutrino mass. (However, it becomes viable if we allow for nonvanishing bare Majorana masses for the left-handed neutrinos.)

To obtain examples beyond the seesaw, consider two sterile neutrinos ($n = 2$).  We wish to solve for complex $M$ with $\operatorname{rank} M = 2$.  Below, we consider the physical setup with $m = 3$; the extension to arbitrary $m$ is obvious.
\begin{itemize}
\item If $\operatorname{rank} B = 0$, then $B = 0$ and $\operatorname{rank} M = 2$ if and only if $\operatorname{rank} D = 2$.
\item If $\operatorname{rank} B = 1$, then up to conjugation by $\operatorname{diag}(I_3, \left[\begin{smallmatrix} 0 & 1 \\ 1 & 0 \end{smallmatrix}\right])$, we can write
\begin{equation}
M = \left[\begin{array}{ccc|cc}
0 & 0 & 0 & a & \alpha a \\
0 & 0 & 0 & b & \alpha b \\
0 & 0 & 0 & c & \alpha c \\ \hline
a & b & c & d_{11} & d_{12} \\
\alpha a & \alpha b & \alpha c & d_{12} & 2\alpha d_{12} - \alpha^2 d_{11}
\end{array}\right]
\label{earliertexture}
\end{equation}
where not all of $a, b, c$ are zero and $d_{11}, d_{12}, \alpha$ are unconstrained.  All entries are generically complex.  If $d_{12}\neq \alpha d_{11}$, then the last two columns are linearly independent and the first three columns lie in the span of the last two.  Otherwise, the last two columns are linearly dependent and the one-dimensional span of the first three columns lies outside the span of the last two.
\item If $\operatorname{rank} B = 2$, then the last two columns are linearly independent regardless of $D$, so the first three columns must lie in the span of the last two.  But the only linear combination of the last two columns with vanishing first three entries is the zero vector, so this is impossible.  Hence we cannot have $\operatorname{rank} B = 2$.
\end{itemize}

It is clear that, in general, we must have $\operatorname{rank} B < n$ to ensure that $\operatorname{rank} M = n$.  To proceed to higher values of $n$, it is convenient to instead solve the eigenvalue conditions \eqref{conditions} and then to promote the resulting real parametrized textures to complex ones.

\subsection{Eigenvalue Conditions}

We now specialize to real mass matrices $M$, which allows us to interpret the eigenvalues directly as physical neutrino masses (up to signs) and to write their vanishing conditions as simple algebraic conditions on the entries.  We leave $m$ arbitrary unless otherwise specified.

\subsubsection{\texorpdfstring{$n = 1$}{n = 1}}

As already mentioned, the case $n = 1$ yields no new textures beyond those corresponding to the ordinary seesaw mechanism, which entails perturbing around the trivial solution $B = 0$ to the conditions \eqref{conditions}.  Indeed, in this case, the only constraint is $G = B^T B = 0$, but if $B$ is real, then $G = 0$ if and only if $B = 0$.

\subsubsection{\texorpdfstring{$n = 2$}{n = 2}}

It is straightforward to classify all solutions for $n = 2$.  Without assuming that $D$ is diagonal, we write
\begin{equation}
D = \left[\begin{array}{cc} d_{11} & d_{12} \\ d_{12} & d_{22} \end{array}\right], \qquad G = B^T B = \left[\begin{array}{cc} g_{11} & g_{12} \\ g_{12} & g_{22} \end{array}\right].
\end{equation}
The characteristic polynomial of $M$ is $x^{m-2}P_4(x)$ where
\begin{equation}
P_4(x) = x^4 - (\Tr D)x^3 + (\det D - \Tr G)x^2 + (d_{11}g_{22} - 2d_{12}g_{12} + d_{22}g_{11})x + \det G,
\label{quartic}
\end{equation}
so we must satisfy two constraints:
\begin{equation}
\det G = 0, \qquad d_{11}g_{22} - 2d_{12}g_{12} + d_{22}g_{11} = 0.
\end{equation}
There are two classes of solutions: either $\operatorname{rank} G = 0$ or $\operatorname{rank} G = 1$.

If $\operatorname{rank} G = 0$, then $G$ and $B$ are identically zero and both constraints are manifestly satisfied; $D$ can be arbitrary.  This is again the case of the ordinary seesaw mechanism.

If $\operatorname{rank} G = 1$, then not all entries of $G$ are zero.  The possible solutions are as follows:
\begin{align}
(D, G) &= \left(\left[\begin{array}{cc} d_{11} & d_{12} \\ d_{12} & 0 \end{array}\right], \left[\begin{array}{cc} g_{11} & 0 \\ 0 & 0 \end{array}\right]\right), \qquad g_{11}\neq 0, \label{n2rank1soln-1} \\
(D, G) &= \left(\left[\begin{array}{cc} 0 & d_{12} \\ d_{12} & d_{22} \end{array}\right], \left[\begin{array}{cc} 0 & 0 \\ 0 & g_{22} \end{array}\right]\right), \qquad g_{22}\neq 0, \label{n2rank1soln-2} \\
(D, G) &= \left(\left[\begin{array}{cc} d_{11} & d_{12} \\ d_{12} & 2d_{12}g_{12}/g_{11} - d_{11}g_{12}^2/g_{11}^2 \end{array}\right], \left[\begin{array}{cc} g_{11} & g_{12} \\ g_{12} & g_{12}^2/g_{11} \end{array}\right]\right), \qquad g_{11}, g_{12}\neq 0, \label{n2rank1soln-3}
\end{align}
with the $d_{ij}$ arbitrary in all cases.  The first two classes of solutions are related by conjugation by $\left[\begin{smallmatrix} 0 & 1 \\ 1 & 0 \end{smallmatrix}\right]$.  The columns of $B$ are constrained by the entries of $G$.  Namely, since $\operatorname{rank} B = 1$, the three respective cases correspond to:
\begin{alignat}{2}
B &= \left[\begin{array}{cc} \vec{b} & 0 \end{array}\right], \qquad & g_{11} &= |\vec{b}|^2, \\
B &= \left[\begin{array}{cc} 0 & \vec{b} \end{array}\right], \qquad & g_{22} &= |\vec{b}|^2, \\
B &= \left[\begin{array}{cc} \vec{b} & \alpha\vec{b} \end{array}\right], \qquad & g_{11} &= |\vec{b}|^2, \qquad g_{12} = \alpha|\vec{b}|^2,
\end{alignat}
with $\vec{b}\neq 0$ and $\alpha\neq 0$.

One can perturb the above solutions to generate two small but nonzero eigenvalues for $M$.  All nontrivial eigenvalues arise as the roots of the quartic polynomial \eqref{quartic}.  One way to obtain an analytically tractable model is to consider perturbations that turn this quartic polynomial into a quadratic polynomial in $x^2$.  For example, one can set $D = 0$ and then perturb only $G$ (i.e., $B$).  In this case, the characteristic polynomial of $M$ is
\begin{equation}
x^{m-2}(x^4 - (\Tr G)x^2 + \det G),
\end{equation}
whose roots are
\begin{equation}
x = \underbrace{0, \ldots, 0}_{m-2}, \pm\sqrt{\frac{\Tr G\pm' \sqrt{(\Tr G)^2 - 4\det G}}{2}}
\end{equation}
for all four choices of signs $\pm, \pm'$.  However, this model is not phenomenologically viable, as it does not produce the experimentally required mass splitting (the smallest two eigenvalues are always $\pm$ of each other).

We now present some explicit textures for the physically relevant case of $m = 3$, i.e., $(m, n) = (3, 2)$ (the extension to arbitrary $m$ is obvious).  The only potential novelty arises when $\operatorname{rank} G = 1$.  The textures
\begin{equation}
\left[\begin{array}{ccc|cc}
0 & 0 & 0 & a & 0 \\
0 & 0 & 0 & b & 0 \\
0 & 0 & 0 & c & 0 \\ \hline
a & b & c & \lambda & \mu \\
0 & 0 & 0 & \mu & 0
\end{array}\right], \qquad
\left[\begin{array}{ccc|cc}
0 & 0 & 0 & 0 & a \\
0 & 0 & 0 & 0 & b \\
0 & 0 & 0 & 0 & c \\ \hline
0 & 0 & 0 & 0 & \mu \\
a & b & c & \mu & \lambda
\end{array}\right], \qquad
\left[\begin{array}{ccc|cc}
0 & 0 & 0 & a & \alpha a \\
0 & 0 & 0 & b & \alpha b \\
0 & 0 & 0 & c & \alpha c \\ \hline
a & b & c & \lambda & \mu \\
\alpha a & \alpha b & \alpha c & \mu & 2\alpha\mu - \alpha^2\lambda
\end{array}\right]
\label{mn32textures}
\end{equation}
parametrize the solutions \eqref{n2rank1soln-1}, \eqref{n2rank1soln-2}, and \eqref{n2rank1soln-3}, respectively.  The only conditions on the parameters are that $a^2 \linebreak[1] + \linebreak[1] b^2 \linebreak[1] + \linebreak[1] c^2\neq 0$ and $\alpha\neq 0$.  The first two textures, which are equivalent by a change of basis, are also examples of the cross texture \eqref{crosstexture}.  As expected, the last texture coincides with \eqref{earliertexture}. (See \cite{Babu:2022non} for an application of the case $\lambda = 0$ and $\alpha = 0$.) Perturbing these textures, such as by taking
\begin{equation}
\left[\begin{array}{ccc|cc}
0 & 0 & 0 & a & 0 \\
0 & 0 & 0 & b & 0 \\
0 & 0 & 0 & c & 0 \\ \hline
a & b & c & \lambda & \mu \\
0 & 0 & 0 & \mu & 0
\end{array}\right] \to
\left[\begin{array}{ccc|cc}
0 & 0 & 0 & a & \epsilon_1 \\
0 & 0 & 0 & b & \epsilon_2 \\
0 & 0 & 0 & c & \epsilon_3 \\ \hline
a & b & c & \lambda & \mu \\
\epsilon_1 & \epsilon_2 & \epsilon_3 & \mu & \epsilon
\end{array}\right], \qquad
\left[\begin{array}{ccc|cc}
0 & 0 & 0 & 0 & a \\
0 & 0 & 0 & 0 & b \\
0 & 0 & 0 & 0 & c \\ \hline
0 & 0 & 0 & 0 & \mu \\
a & b & c & \mu & \lambda
\end{array}\right] \to
\left[\begin{array}{ccc|cc}
0 & 0 & 0 & \epsilon_1 & a \\
0 & 0 & 0 & \epsilon_2 & b \\
0 & 0 & 0 & \epsilon_3 & c \\ \hline
\epsilon_1 & \epsilon_2 & \epsilon_3 & \epsilon & \mu \\
a & b & c & \mu & \lambda
\end{array}\right],
\end{equation}
gives one vanishing mass, two small masses, and two large masses.  These are all of the $(m, n) \linebreak[1] = \linebreak[1] (3, 2)$ textures beyond the seesaw.

Let us make some brief comments about mixing parameters.  To obtain realistic models, we must generate not only small neutrino masses, but also small mixing angles.  To respect constraints from experiment, oscillations between light and heavy states cannot be too large.  While our approach guarantees small masses, the smallness of the resulting mixing angles imposes constraints on the parametrized textures that must be determined by further analysis.  One complication is that, in general, the eigenvectors of a matrix are not continuous functions of its entries, so the mixing matrix cannot always be reliably computed before adding the perturbation to the mass matrix $M$ \cite{Kato}.

Again, we consider a simple example with $n = 2$.  For simplicity, we assume that $M$ is real, so that it suffices to compute eigenvectors rather than singular vectors.  We also assume that the desired perturbation will not change the mixing matrix drastically, which allows us to derive the approximate mixing matrix analytically before adding the perturbation to $M$.  We show that the off-diagonal block (the $\nu N$ mixing matrix) can be made parametrically small.

Specifically, consider \eqref{mn32textures}.  The smallness of the mixing parameters is controlled by the relative size of $a, b, c$ and $\mu$, so for simplicity, we set $\lambda = 0$ and $\alpha = 0$.  The corresponding texture and its eigenvalues are
\begin{equation}
M = \left[\begin{array}{ccc|cc}
0 & 0 & 0 & 0 & a \\
0 & 0 & 0 & 0 & b \\
0 & 0 & 0 & 0 & c \\ \hline
0 & 0 & 0 & 0 & \mu \\
a & b & c & \mu & 0
\end{array}\right], \qquad
\operatorname{spec}(M) = \left\{0, 0, 0, \lambda_\pm\equiv \pm\sqrt{\vec{b}^2 + \mu^2}\right\}, \qquad
\vec{b}\equiv \left[\begin{array}{c} a \\ b \\ c \end{array}\right].
\end{equation}
The singular values are obtained from the eigenvalues by taking $|\lambda_-| = -\lambda_-$.  The normalized eigenvectors corresponding to $\lambda_-$ and $\lambda_+$ are, respectively,
\begin{equation}
\vec{v}_-\equiv \frac{1}{\sqrt{2(\vec{b}^2 + \mu^2)}}\left[\begin{array}{c}
-\vec{b} \\ \hline
-\mu \\
\sqrt{\vec{b}^2 + \mu^2}
\end{array}\right], \qquad
\vec{v}_+\equiv \frac{1}{\sqrt{2(\vec{b}^2 + \mu^2)}}\left[\begin{array}{c}
\vec{b} \\ \hline
\mu \\
\sqrt{\vec{b}^2 + \mu^2}
\end{array}\right].
\end{equation}
The orthogonal matrix that diagonalizes $M$ is given by
\begin{equation}
O = \left[\begin{array}{c|c|c|c|c} \ast & \ast & \ast & \vec{v}_- & \vec{v}_+ \end{array}\right]
\end{equation}
where the $\ast$ columns comprise an orthonormal basis for the degenerate 0 subspace.  Up to phases, the observable off-diagonal mixing parameters are the first three components of $\vec{v}_\pm$.  By taking $\mu$ sufficiently large relative to $a$, $b$, $c$, we can make all of these components arbitrarily small.  In this limit, the lower-left block of $O$ also becomes arbitrarily small, so that the PMNS matrix becomes arbitrarily close to unitary.  Note that even though this mechanism requires a hierarchy, this is not a seesaw mechanism: in the seesaw, the smallness of masses and mixings is correlated as in \eqref{seesawmixing}.  Here, small masses are already guaranteed by the structure of the texture itself.

\subsubsection{\texorpdfstring{$n\geq 3$}{n >= 3}}

Now consider arbitrary $n$.  The simplest class of solutions beyond the seesaw is that for which $\operatorname{rank} G = 1$; we focus on these solutions.  If $\operatorname{rank} G = 1$, then $\operatorname{rank} B = 1$ and we can write
\begin{equation}
B = uv^T, \qquad G = (u^T u)(vv^T),
\end{equation}
where $u$ and $v$ are nonzero column vectors of length $m$ and $n$, respectively. (One can carry out a similar analysis for higher rank using the singular value decomposition of $B$.)

First suppose (for simplicity, and without loss of generality) that $D$ is diagonal.  The conditions \eqref{conditions} for $M$ to have $m$ vanishing eigenvalues reduce to
\begin{equation}
\sum_{i=1}^n G_{ii}\prod_{j\neq i} d_j = 0
\end{equation}
because all minors of $G$ aside from the $1\times 1$ minors vanish, so the only nontrivial condition in \eqref{conditions} comes from $d = 1$ and $r = n - 1$.  This is equivalent to
\begin{equation}
\sum_{i=1}^n v_i^2\prod_{j\neq i} d_j = 0.
\label{diequation}
\end{equation}
If two or more of the $d_i$ vanish, then this condition is automatically satisfied.  If exactly one of the $d_i$ vanishes, say $d_{i_0}$, then this condition reduces to $v_{i_0} = 0$.  The remaining case is that all of the $d_i$ are nonzero.  To simplify the problem, suppose that we have
\begin{equation}
\sum_{i=1}^n a_i v_i^2 = 0
\label{aiequation}
\end{equation}
with all $a_i$ nonzero (clearly, the $a_i$ cannot all be positive).  Then:
\begin{itemize}
\item For $n$ even, nonzero solutions $d_i$ to \eqref{diequation} are in one-to-one correspondence with nonzero solutions $a_i$ to \eqref{aiequation}.
\item For $n$ odd, nonzero solutions $d_i$ to \eqref{diequation} are in two-to-one correspondence with nonzero solutions $a_i$ to \eqref{aiequation} satisfying $a_1\cdots a_n > 0$.
\end{itemize}
Indeed, to solve $a_i = \prod_{j\neq i} d_j$ for the $d_i$, write $a_1\cdots a_n = (d_1\cdots d_n)^{n-1}$.  If $n$ is even, then there exists a unique solution with all $d_i$ real:
\begin{equation}
\operatorname{sgn}(a_1\cdots a_n)|a_1\cdots a_n|^{\frac{1}{n-1}} = d_1\cdots d_n
\end{equation}
and hence
\begin{equation}
d_i = \frac{\operatorname{sgn}(a_1\cdots a_n)|a_1\cdots a_n|^{\frac{1}{n-1}}}{a_i} = \frac{\operatorname{sgn}\big(\prod_{j\neq i} a_j\big)\big|\prod_{j\neq i} a_j\big|^{\frac{1}{n-1}}}{|a_i|^{\frac{n - 2}{n - 1}}}.
\end{equation}
If $n$ is odd, then there exists no solution with all $d_i$ real when $a_1\cdots a_n < 0$, but there exist two solutions when $a_1\cdots a_n > 0$:
\begin{equation}
\pm(a_1\cdots a_n)^{\frac{1}{n-1}} = d_1\cdots d_n
\end{equation}
and hence
\begin{equation}
d_i = \pm\frac{(a_1\cdots a_n)^{\frac{1}{n-1}}}{a_i} = \pm\frac{\operatorname{sgn}(a_i)\big|\prod_{j\neq i} a_j\big|^{\frac{1}{n-1}}}{|a_i|^{\frac{n - 2}{n - 1}}} = \pm\frac{\operatorname{sgn}\big(\prod_{j\neq i} a_j\big)\big|\prod_{j\neq i} a_j\big|^{\frac{1}{n-1}}}{|a_i|^{\frac{n - 2}{n - 1}}}.
\end{equation}
To summarize, we have found solutions to \eqref{conditions} of the form
\begin{equation}
M = \left[\begin{array}{cc} 0 & uv^T \\ vu^T & D \end{array}\right]
\end{equation}
where $u, v$ are nonzero, $D = \operatorname{diag}(d_1, \ldots, d_n)$, and one of the following holds:
\begin{itemize}
\item $\operatorname{rank} D\leq n - 2$,
\item $\operatorname{rank} D = n - 1$ with $d_i = v_i = 0$ for some $i$,
\item $\operatorname{rank} D = n$ with $\sum_{i=1}^n v_i^2\prod_{j\neq i} d_j = 0$.
\end{itemize}
The most general solution with $\operatorname{rank} G = \operatorname{rank} B = 1$ is obtained by a change of basis:
\begin{equation}
M = \left[\begin{array}{cc} 0 & u(Ov)^T \\ (Ov)u^T & ODO^T \end{array}\right],
\end{equation}
where $O$ is an orthogonal matrix and $D' = ODO^T$ is a real symmetric matrix.

As a special but useful situation in which $\operatorname{rank} G = 1$, consider the case where $B$ has a single nonvanishing column:
\begin{equation}
B = \left[\begin{array}{c|c|c|c|c|c|c} \vec{0} & \cdots & \vec{0} & \vec{b} & \vec{0} & \cdots & \vec{0} \end{array}\right].
\end{equation}
Then $G$ has a single nonzero entry and, without assuming that $D$ is diagonal, the only requirement that needs to be imposed for $M$ to have $m$ vanishing eigenvalues is the vanishing of a single $(n - 1)\times (n - 1)$ minor of $D$ (we provide a proof of this statement in Appendix \ref{diagonalgram}).  For example, with $(m, n) = (3, 3)$, we can take
\begin{equation}
M = \left[\begin{array}{ccc|ccc}
0 & 0 & 0 & a & 0 & 0 \\
0 & 0 & 0 & b & 0 & 0 \\
0 & 0 & 0 & c & 0 & 0 \\ \hline
a & b & c & d_{11} & d_{12} & d_{13} \\
0 & 0 & 0 & d_{12} & d_{22} & d_{23} \\
0 & 0 & 0 & d_{13} & d_{23} & d_{33}
\end{array}\right]
\end{equation}
where $a^2 + b^2 + c^2\neq 0$.  Then the only requirement for obtaining three zero eigenvalues is that the lower-right $2\times 2$ minor of $D$ should vanish: $d_{23}^2 - d_{22}d_{33} = 0$.  So we obtain a $6\times 6$ example without a hierarchy between Dirac and Majorana masses.

Let us again comment on mixing parameters, this time in the context of an example with $(m, n) = (3, 3)$.  The seed texture corresponding to \eqref{exampletexture} in the main text takes the form
\begin{equation}
M = \left[\begin{array}{ccc|ccc}
0 & 0 & 0 & 0 & 0 & 0 \\
0 & 0 & 0 & 0 & x & 0 \\
0 & 0 & 0 & 0 & 0 & 0 \\ \hline
0 & 0 & 0 & & & \\
0 & x & 0 & & D & \\
0 & 0 & 0 & & &
\end{array}\right].
\end{equation}
To obtain three vanishing eigenvalues irrespective of the value of $x$, we choose the entries $d_{ij}$ of $D$ to satisfy $d_{11}d_{33} \linebreak[1] - \linebreak[1] d_{13}^2 = 0$.  To obtain small mixing parameters (subject to the same caveats and assumptions as in our discussion of the $n = 2$ case), it is convenient to suppose that the matrix $D$ is characterized by a scale $d$ such that taking $x\ll d$ allows one to compute the mixing parameters in the seesaw approximation (independently of the masses, which are generated by perturbations to the seed texture).  For this purpose, one must avoid making pathological choices such as
\begin{equation}
D = \left[\begin{array}{ccc} d & d & d \\ d & d & d \\ d & d & d \end{array}\right],
\label{pathological}
\end{equation}
for which the mixing parameters remain $O(1)$ regardless of how small $x$ is taken relative to $d$.  The problem with \eqref{pathological} is that the matrix $D$ is singular, which has the consequence that the ostensibly ``heavy'' Majorana degrees of freedom contain a massless mode and are therefore impossible to decouple from the ``light'' degrees of freedom by making the Dirac mass $x$ arbitrarily small. (One sees directly that the nonexistence of $D^{-1}$ invalidates the seesaw approximation \eqref{seesawmixing}.) On the other hand, it is possible to choose invertible $D$, all of whose entries have absolute values of order $d$, satisfying the vanishing condition on the appropriate $2\times 2$ minor.  Examples include
\begin{equation}
D = \left[\begin{array}{ccc} d & \pm d & d \\ \pm d & d & \mp d \\ d & \mp d & d \end{array}\right] \Longleftrightarrow D^{-1} = \left[\begin{array}{ccc} 0 & \pm 1/2d & 1/2d \\ \pm 1/2d & 0 & \mp 1/2d \\ 1/2d & \mp 1/2d & 0 \end{array}\right].
\end{equation}
The corresponding inverses have entries with absolute values of order $1/d$, allowing one to use the seesaw approximation for the mixing parameters.

Finally, although we have so far considered real symmetric $M$, note that the general solution to $\operatorname{rank} M = n$ for complex symmetric $M$ can easily be bootstrapped from the real solution to the eigenvalue conditions \eqref{conditions}.  In particular, the most general complex symmetric $M$ with $\operatorname{rank} B = 1$ and $D$ diagonal that satisfies $\operatorname{rank} M = n$ takes the form
\begin{equation}
M = \left[\begin{array}{cc} 0 & uv^T \\ vu^T & D \end{array}\right], \qquad \sum_{i=1}^n v_i^2\prod_{j\neq i} d_j = 0,
\end{equation}
where $u$ and $v$ are nonzero column vectors of length $m$ and $n$, respectively, and $D = \operatorname{diag}(d_1, \ldots, d_n)$.  All parameters are now complex.  Lifting the assumption that $D$ is diagonal, the most general solution with $\operatorname{rank} B = 1$ is obtained by a change of basis:
\begin{equation}
M = \left[\begin{array}{cc} 0 & u(Uv)^T \\ (Uv)u^T & UDU^T \end{array}\right],
\end{equation}
where $U$ is a unitary matrix.

\section{Derivations}

\subsection{Mass Matrix with No Structure}

Let $M$ be an $N\times N$ matrix.  When does $M$ have (at least) $m$ vanishing eigenvalues?  To answer this question, we use that for $n\times n$ matrices $A$ and $B$ \cite{marcusdetsum},
\begin{equation}
\det(A + B) = \sum_{r=0}^n \sum_{\alpha, \beta} (-1)^{s(\alpha) + s(\beta)}\det(A[\alpha|\beta])\det(B(\alpha|\beta))
\label{determinantsum}
\end{equation}
where $\alpha, \beta$ are strictly increasing integer sequences of length $r$ chosen from $1, \ldots, n$; $A[\alpha|\beta]$ is the $r\times r$ submatrix of $A$ corresponding to rows $\alpha$ and columns $\beta$; $B(\alpha|\beta)$ is the $(n - r)\times (n - r)$ submatrix of $B$ corresponding to rows complementary to $\alpha$ and columns complementary to $\beta$; and $s(\alpha)$ is the sum of the integers in $\alpha$.  When $A$ is diagonal, \eqref{determinantsum} simplifies to
\begin{equation}
\det(A + B) = \sum_{r=0}^n \sum_{|\alpha| = r} \det(A[\alpha|\alpha])\det(B(\alpha|\alpha)).
\label{diagsum}
\end{equation}
The formula \eqref{diagsum} implies the following standard formula for the coefficients of the characteristic polynomial of $M$: $\det(xI - M) = x^N + c_{N-1}x^{N-1} + \cdots + c_1 x + c_0$ where
\begin{equation}
c_r = (-1)^{N-r}\sum_{|\alpha| = r} \det(M(\alpha|\alpha)).
\end{equation}
For $M$ to have $m$ vanishing eigenvalues, we must have $c_0 = \cdots = c_{m-1} = 0$, so the sum of all principal $(N - r)\times (N - r)$ minors of $M$ must vanish for each $r = 0, \ldots, m - 1$.  If $M$ has rank at most $N - m$, then this is clearly the case.

\subsection{Mass Matrix with Block Structure}

The above general conditions for the vanishing of $m$ eigenvalues do not assume any structure on $M$.  We are interested in the case \eqref{massmatrix}, where $M$ is a complex symmetric matrix whose upper-left $m\times m$ block vanishes:
\begin{equation}
M = \left[\begin{array}{cc} \mathbf{0}_{m\times m} & B \\ B^T & D \end{array}\right].
\end{equation}
Here, $B$ is an $m\times n$ complex matrix, $D$ is an $n\times n$ complex symmetric matrix, and $N = m + n$.

To determine the most general conditions under which such an $M$ has $m$ vanishing eigenvalues, we first write the characteristic polynomial of $M$ as\footnote{We have used that the determinant of a block matrix $M = \left[\begin{smallmatrix} A & B \\ C & D \end{smallmatrix}\right]$, where $A$ and $D$ are square matrices with $A$ invertible, is $\det(M) \linebreak[1] = \linebreak[1] \det(A)\det(D - CA^{-1}B)$.}
\begin{equation}
\left|\begin{array}{cc} x\mathbf{1}_{m\times m} & -B \\ -B^T & x\mathbf{1}_{n\times n} - D \end{array}\right| = x^{m-n}P_{2n}(x), \qquad P_{2n}(x)\equiv \det(x^2\mathbf{1}_{n\times n} - xD - G),
\end{equation}
where $G$ is the Gram matrix of column vectors of $B$ with respect to the \emph{real} inner product:
\begin{equation}
G = B^T B.
\end{equation}
For $M$ to have $m$ zero eigenvalues, $P_{2n}(x)$ must contain no terms of degree $0, \ldots, n - 1$ in $x$.  Since the rank of $G$ is no greater than $m$,\footnote{The rank of $G$ does not follow straightforwardly from the rank of $B$.  For real matrices $B$, we have $\operatorname{rank} B^T B = \operatorname{rank} BB^T = \operatorname{rank} B$; for complex $B$, an analogous statement holds with $B^\dag$ in the place of $B^T$.} we need only demand the vanishing of the coefficients of $x^{n - 1}, \ldots, x^{n - \min(m, n)}$ in $P_{2n}(x)$.  These $\min(m, n)$ conditions are equivalent to $M$ having at least $m$ vanishing eigenvalues.

To write these conditions explicitly, we make the following simplification.  Since there exists a unitary matrix $U$ such that $U^T DU$ is diagonal with nonnegative real entries, without loss of generality, we may work in a basis where the lower-right block of $M$ is diagonal---if necessary, by changing basis purely within the space of sterile neutrinos as follows:
\begin{equation}
M\to \left[\begin{array}{cc} \mathbf{1} & \mathbf{0} \\ \mathbf{0} & U^T \end{array}\right]\left[\begin{array}{cc} \mathbf{0} & B \\ B^T & D \end{array}\right]\left[\begin{array}{cc} \mathbf{1} & \mathbf{0} \\ \mathbf{0} & U \end{array}\right] = \left[\begin{array}{cc} \mathbf{0} & BU \\ U^T B^T & U^T DU \end{array}\right].
\end{equation}
So let $D$ be diagonal with entries $d_i\geq 0$.  Using \eqref{diagsum}, we can write
\begin{align}
P_{2n}(x) &= \sum_{r=0}^n (-1)^{n-r}x^r\sum_{|\alpha| = r} (x - d_{\alpha(1)})\cdots (x - d_{\alpha(r)})\det(G(\alpha|\alpha)) \\
&= (-1)^n\sum_{r=0}^n \sum_{p=0}^r (-1)^{r - p}\sum_{|\alpha| = r} e_p(d_{\alpha(1)}, \ldots, d_{\alpha(r)})\det(G(\alpha|\alpha))x^{2r-p},
\end{align}
where the $e_p$ are elementary symmetric polynomials in $r$ variables.  By using
\begin{equation}
\sum_{r=0}^n \sum_{p=0}^r C(r, p)x^{2r - p} = \sum_{k=0}^{2n} x^{2n - k}\sum_{\ell=0}^{\lfloor k/2\rfloor} C(n - \ell, k - 2\ell)
\end{equation}
to rearrange the sum, we can further write
\begin{equation}
P_{2n}(x) = \sum_{k=0}^{2n} x^{2n - k}\sum_{\ell=0}^{\lfloor k/2\rfloor} (-1)^{k - \ell}\sum_{|\alpha| = n - \ell} e_{k - 2\ell}(d_{\alpha(1)}, \ldots, d_{\alpha(n - \ell)})\det(G(\alpha|\alpha)).
\end{equation}
So for each $k = n + 1, \ldots, 2n$, we demand that
\begin{equation}
\sum_{\ell = k - n}^{\lfloor k/2\rfloor} (-1)^{k - \ell}\sum_{|\alpha| = n - \ell} e_{k - 2\ell}(d_{\alpha(1)}, \ldots, d_{\alpha(n - \ell)})\det(G(\alpha|\alpha)) = 0,
\end{equation}
where we have changed the lower limit of summation on $\ell$ from 0 to $k - n$ because below this range, the elementary symmetric polynomial $e_{k - 2\ell}$ in $n - \ell$ variables vanishes.  Setting $d = k - n$ and $r = n - \ell$, and ignoring an overall sign, we can equivalently write these conditions as follows: for each $d = 1, \ldots, n$,
\begin{equation}
\sum_{r = \lceil\frac{n - d}{2}\rceil}^{n - d} (-1)^r\sum_{|\alpha| = r} e_{2r - n + d}(d_{\alpha(1)}, \ldots, d_{\alpha(r)})\det(G(\alpha|\alpha)) = 0.
\end{equation}
This is precisely \eqref{conditions}.  Note that since $G(\alpha|\alpha)$ is at least a $d\times d$ matrix and the rank of $G$ is at most $m$, we have $\det(G(\alpha|\alpha)) = 0$ when $d > m$.  So these conditions are automatically satisfied when $d > m$, meaning that these $n$ conditions really reduce to $\min(m, n)$ conditions.  Similarly, we could restrict the range of summation on $r$ to start at $r = \max(n - m, \lceil\frac{n - d}{2}\rceil)$.

\subsection{Diagonal Gram Matrix} \label{diagonalgram}

Let us now assume that $G$, rather than $D$, is diagonal with entries $g_i$.  Using \eqref{diagsum}, we can write
\begin{align}
P_{2n}(x) &= \sum_{r=0}^n (-x)^{n-r}\sum_{|\alpha| = r} (x^2 - g_{\alpha(1)})\cdots (x^2 - g_{\alpha(r)})\det(D(\alpha|\alpha)) \\
&= (-1)^n\sum_{r=0}^n \sum_{p=0}^r (-1)^{r - p}\sum_{|\alpha| = r} e_p(g_{\alpha(1)}, \ldots, g_{\alpha(r)})\det(D(\alpha|\alpha))x^{n + r - 2p}.
\end{align}
Further using
\begin{equation}
\sum_{r=0}^n \sum_{p=0}^r C(r, p)x^{n + r - 2p} = \sum_{k=-n}^n x^{n+k}\sum_{\ell=0}^{\left\lfloor\frac{n - |k|}{2}\right\rfloor} C\left(|k| + 2\ell, \frac{|k| - k}{2} + \ell\right)
\end{equation}
gives
\begin{equation}
P_{2n}(x) = (-1)^n\sum_{k=-n}^n x^{n+k}\sum_{\ell=0}^{\left\lfloor\frac{n - |k|}{2}\right\rfloor} (-1)^{(|k| + k)/2 + \ell}\sum_{|\alpha| = |k| + 2\ell} e_{(|k| - k)/2 + \ell}(g_{\alpha(1)}, \ldots, g_{\alpha(|k| + 2\ell)})\det(D(\alpha|\alpha)).
\end{equation}
We want $P_{2n}$ to have no terms of degree $0, \ldots, n - 1$.  So we demand that
\begin{equation}
\sum_{\ell=0}^{\left\lfloor\frac{n - k}{2}\right\rfloor} (-1)^\ell\sum_{|\alpha| = k + 2\ell} e_{k + \ell}(g_{\alpha(1)}, \ldots, g_{\alpha(k + 2\ell)})\det(D(\alpha|\alpha)) = 0
\end{equation}
for $k = 1, \ldots, n$.  If only a single diagonal entry $g_{i_0}$ is nonzero, then $e_p = 0$ for $p > 1$, so the only nontrivial condition corresponds to $k = 1$ and $\ell = 0$:
\begin{equation}
\sum_{i=1}^n g_i\det(D(i|i)) = 0 \Longleftrightarrow \det(D(i_0|i_0)) = 0.
\label{singleentrycond}
\end{equation}

\subsection{Rank-One Gram Matrix}

Now suppose that $\operatorname{rank} G = 1$.  First note that if $A$ and $B$ are $n\times n$ matrices with $\operatorname{rank} A = 1$, then
\begin{equation}
\det(A + B) = \det B + \sum_{\alpha, \beta = 1}^n (-1)^{\alpha + \beta}A_{\alpha\beta}\det(B(\alpha|\beta)),
\label{rankonedet}
\end{equation}
which follows from keeping only the $r = 0, 1$ terms in \eqref{determinantsum}.  Letting $E\equiv x\mathbf{1}_{n\times n} - D$, we have from \eqref{rankonedet} that
\begin{equation}
P_{2n}(x) = \det(xE - G) = x^n\det E - x^{n-1}\sum_{\alpha, \beta = 1}^n (-1)^{\alpha + \beta}G_{\alpha\beta}\det(E(\alpha|\beta)).
\end{equation}
Since $P_{2n}$ has no terms of degree lower than $n - 1$ in $x$, it suffices to demand the vanishing of the $x^{n-1}$ term.  The coefficient of this term is
\begin{equation}
[x^{n-1}]P_{2n}(x) = -\sum_{\alpha, \beta = 1}^n (-1)^{\alpha + \beta}G_{\alpha\beta}\det(E|_{x=0}(\alpha|\beta)) = (-1)^n\sum_{\alpha, \beta = 1}^n (-1)^{\alpha + \beta}G_{\alpha\beta}\det(D(\alpha|\beta)).
\end{equation}
Writing $G = (u^T u)(vv^T)$, the vanishing of this coefficient is equivalent to
\begin{equation}
\sum_{i=1}^n v_i^2\det(D(i|i)) + 2\sum_{i < j} (-1)^{i + j}v_i v_j\det(D(i|j)) = 0.
\label{rankonecond}
\end{equation}
If $D$ is diagonal, then the condition \eqref{rankonecond} reduces to \eqref{diequation}.  If $G$ is diagonal (which, since $\operatorname{rank} G = 1$, implies that $G$ has a single nonzero entry), then this condition reduces to \eqref{singleentrycond}.

\subsection{Comments on Vanishing Gram Matrix}

Here, we comment on a mathematical generalization of the seesaw mechanism that yields small nonzero eigenvalues for perturbations of \emph{complex} symmetric mass matrices $M$.  However, since these comments concern the eigenvalues rather than the singular values of complex $M$, they are not relevant for physical neutrino masses.

The observation is simple: if $G = 0$, then the vanishing conditions \eqref{conditions} are manifestly sat\-is\-fied.  Therefore, we can generalize the seesaw texture \eqref{seesawtexture} by perturbing matrices $B$ that satisfy $B^T B = 0$ rather than merely $B = 0$.  Any such $B$ that is nonzero must be complex (otherwise, we would have $0 = \operatorname{rank} B^T B = \operatorname{rank} B$ and hence $B = 0$).  The columns of $B$ then have zero length and are mutually orthogonal with respect to the real inner product.

This is a generalization of the seesaw mechanism in that the lower-right block $D$ is completely arbitrary; no conditions involving $D$ need to be imposed to satisfy \eqref{conditions}.

\subsubsection{\texorpdfstring{$n = 1$}{n = 1}}

If $n = 1$, then this class of examples is the most general solution because the only constraint is $B^T B = 0$; $D$ can be arbitrary.  To satisfy this constraint nontrivially (i.e., with $B$ not identically zero), $B$ must have at least two nonzero components, and not all components of $B$ can be real.

Writing $D = [d]$ and $B = (b_1, \ldots, b_m)^T$, the characteristic polynomial of $M$ is
\begin{equation}
x^{m-1}\left(x^2 - dx - \sum_{i=1}^m b_i^2\right),
\end{equation}
whose roots are
\begin{equation}
x = \underbrace{0, \ldots, 0}_{m-1}, \frac{d\pm \sqrt{d^2 + 4\sum_{i=1}^m b_i^2}}{2}.
\end{equation}
If $B = (\beta_1, \ldots, \beta_m)^T$ where $\beta_1^2 + \cdots + \beta_m^2 = 0$, then the roots are
\begin{equation}
x = \underbrace{0, \ldots, 0}_m, d.
\end{equation}
If $B = (\beta_1 + \epsilon_1, \ldots, \beta_m + \epsilon_m)^T$ where $\beta_1^2 + \cdots + \beta_m^2 = 0$, then the roots are
\begin{equation}
x = \underbrace{0, \ldots, 0}_{m-1}, \frac{d\pm \sqrt{d^2 + 4\sum_{i=1}^m (2\beta_i\epsilon_i + \epsilon_i^2)}}{2}.
\end{equation}
We assume no hierarchy between $\beta$ and $d$, but rather that they are of the same order of magnitude.  If $\epsilon\ll \beta, d$, then the roots become
\begin{equation}
x = \underbrace{0, \ldots, 0}_{m-1}, -\frac{2}{d}\sum_{i=1}^m \beta_i\epsilon_i + O(\epsilon^2), d + O(\epsilon).
\end{equation}
Unlike in the ordinary seesaw mechanism (with $\beta = 0$), the lightest nonzero eigenvalue is only suppressed by a single power of $\epsilon$ rather than two powers.  This feature of the light eigenvalues persists at higher $n$ (see Footnote \ref{firstorder}).

Note that the eigenvector of
\begin{equation}
M = \left[\begin{array}{cc} 0 & \smash{\vec{\beta}} \\ \smash{\vec{\beta}^T} & d \end{array}\right], \qquad \vec{\beta}^2 = 0
\end{equation}
with eigenvalue $d$ is $\left[\begin{smallmatrix} \smash{\vec{\beta}} \vphantom{0} \\ d \end{smallmatrix}\right]$, while the eigenvectors with eigenvalue 0 take the form
\begin{equation}
\left[\begin{array}{c} \vec{v} \\ 0 \end{array}\right], \qquad \vec{\beta}\cdot\vec{v} = 0,
\end{equation}
since $\vec{\beta}$ is not identically 0.  We see that there exist only $m - 1$ linearly independent eigenvectors with eigenvalue 0.  Hence $M$ is not diagonalizable.

\subsubsection{\texorpdfstring{$n = 2$}{n = 2}}

In a basis where $D$ is diagonal, the characteristic polynomial of $M$ is $x^{m-2}P_4(x)$ where
\begin{equation}
P_4(x) = x^4 - (d_{11} + d_{22})x^3 + (d_{11}d_{22} - \Tr G)x^2 + (d_{11}g_{22} + d_{22}g_{11})x + \det G.
\label{quarticdiag}
\end{equation}
We consider perturbations of solutions to $G = 0$.  Suppose we have four complex roots with $|A_1|, |A_2|\gg |a_1|, |a_2|$ and consider the polynomial
\begin{gather}
(x - A_1)(x - A_2)(x - a_1)(x - a_2) \nonumber \\
= x^4 - (A_1 + A_2 + O(a))x^3 + (A_1 A_2 + O(Aa))x^2 - (A_1 A_2(a_1 + a_2) + O(Aa^2))x + A_1 A_2 a_1 a_2.
\end{gather}
Comparing to \eqref{quarticdiag}, we wish to solve
\begin{align}
A_1 + A_2 + O(a) &= d_{11} + d_{22}, \label{firstequations1} \\
A_1 A_2 + O(Aa) &= d_{11}d_{22} - \Tr G, \label{firstequations2} \\
A_1 A_2(a_1 + a_2) + O(Aa^2) &= -d_{11}g_{22} - d_{22}g_{11}, \label{lastequations1} \\
A_1 A_2 a_1 a_2 &= \det G. \label{lastequations2}
\end{align}
It is consistent to assume that the entries of $D$ are of order $A$ (with errors of order $a$) and the entries of $G$ are of order $Aa$ (with errors of order $a^2$).  The first two equations \eqref{firstequations1}--\eqref{firstequations2} require that
\begin{equation}
(A_1, A_2) = \text{some permutation of $(d_{11} + O(G/D), d_{22} + O(G/D))$}.
\end{equation}
Then $A_1 A_2 = d_{11}d_{22} + O(G)$, and the last two equations \eqref{lastequations1}--\eqref{lastequations2} become
\begin{equation}
a_1 + a_2 = -\frac{g_{11}}{d_{11}} - \frac{g_{22}}{d_{22}} + O(G^2/D^3), \qquad a_1 a_2 = \frac{\det G}{d_{11}d_{22}} + O(G^3/D^4)
\end{equation}
(in other words, the errors are suppressed by an additional factor of $G/D^2$).  So $a_1$ and $a_2$ are the solutions to
\begin{equation}
x^2 + \left(\frac{g_{11}}{d_{11}} + \frac{g_{22}}{d_{22}} + O(G^2/D^3)\right)x + \left(\frac{\det G}{d_{11}d_{22}} + O(G^3/D^4)\right) = 0,
\end{equation}
which are
\begin{equation}
x = \frac{1}{2}\left[-\left(\frac{g_{11}}{d_{11}} + \frac{g_{22}}{d_{22}}\right)\pm \sqrt{\left(\frac{g_{11}}{d_{11}} - \frac{g_{22}}{d_{22}}\right)^2 + \frac{4g_{12}^2}{d_{11}d_{22}}}\right] + O(G^2/D^3).
\end{equation}
We can rewrite this in terms of perturbations of $B$.  The general lesson is that the small eigenvalues $a$ are of order $G/D$, but this does not necessarily mean that they are of order $B^2/D$, as in the ordinary seesaw.

For illustration, let the columns of $B$ be
\begin{equation}
(\beta_1 + \epsilon_1, \ldots, \beta_m + \epsilon_m)^T, \qquad (\beta_1' + \epsilon_1', \ldots, \beta_m' + \epsilon_m')^T
\end{equation}
where
\begin{equation}
\beta_1^2 + \cdots + \beta_m^2 = \beta_1'^2 + \cdots + \beta_m'^2 = \beta_1\beta_1' + \cdots + \beta_m\beta_m' = 0
\end{equation}
and the perturbations $\epsilon$ are assumed small relative to $\beta$.  Then we have
\begin{equation}
G = \left[\begin{array}{cc} 2(\beta_1\epsilon_1 + \cdots \beta_m\epsilon_m) & \beta_1\epsilon_1' + \beta_1'\epsilon_1 + \cdots + \beta_m\epsilon_m' + \beta_m'\epsilon_m \\ \beta_1\epsilon_1' + \beta_1'\epsilon_1 + \cdots + \beta_m\epsilon_m' + \beta_m'\epsilon_m & 2(\beta_1'\epsilon_1' + \cdots \beta_m'\epsilon_m') \end{array}\right] + O(\epsilon^2).
\end{equation}
So $G$ is of order $\beta\epsilon$ rather than $B^2\sim \beta^2$.

\subsubsection{\texorpdfstring{$n\geq 3$}{n >= 3}}

Now consider arbitrary $n$.  We assume that $D$ is diagonal with entries $d_i\geq 0$.  The characteristic polynomial of $M$ can be written as $x^{m-n}P_{2n}(x)$ where
\begin{equation}
P_{2n}(x) = \sum_{k=0}^{2n} x^{2n - k}\sum_{\ell=0}^{\lfloor k/2\rfloor} (-1)^{k - \ell}\sum_{|\alpha| = n - \ell} e_{k - 2\ell}(d_{\alpha(1)}, \ldots, d_{\alpha(n - \ell)})\det(G(\alpha|\alpha)).
\label{p2n}
\end{equation}
Instead of considering the most general $D$ and $G$ for which $P_{2n}$ has $n$ vanishing roots, we again focus on the class of examples where the Gram matrix $G$ is identically zero and $D$ is completely unconstrained.

We want to produce $\min(m, n)$ small eigenvalues by perturbing around $G = 0$.  Suppose for simplicity that $m\geq n$.  If $m < n$, then the reasoning below still carries through: in that case, $P_{2n}(x)$ only has terms $x^{2n}, x^{2n - 1}, \ldots, x^{n - m}$ and so automatically has $n - m$ vanishing roots, so demanding that it have $n$ large and $n$ small roots will automatically produce $m$ small roots.

Regarding $G$ as a perturbation ($G/D^2\ll 1$), for each $k$ in \eqref{p2n}, the term in the coefficient of lowest $\ell$ dominates, and for $e_{k - 2\ell}$ in $n - \ell$ variables to be nonzero, we must have $\ell\geq k - n$.  So if $k\leq n$, then we choose $\ell = 0$; otherwise, we choose $\ell = k - n$.  So we have
\begin{align}
P_{2n}(x) &= \sum_{k=0}^n [(-1)^k e_k(d_1, \ldots, d_n) + O(D^{k-2}G)]x^{2n - k} \nonumber \\
&\phantom{==} + \sum_{k=n+1}^{2n} \left[(-1)^n\sum_{|\alpha| = 2n - k} d_{\alpha(1)}\cdots d_{\alpha(2n - k)}\det(G(\alpha|\alpha)) + O(D^{2n - k - 2}G^{k - n + 1})\right]x^{2n - k}.
\end{align}
Now consider
\begin{gather}
(x - A_1)\cdots (x - A_n)(x - a_1)\cdots (x - a_n) = \sum_{k=0}^n ((-1)^k e_k(A_1, \ldots, A_n) + O(A^{k-1}a))x^{2n - k} \nonumber \\
{} + \sum_{k=n+1}^{2n} ((-1)^k A_1\cdots A_n e_{k-n}(a_1, \ldots, a_n) + O(A^{n - 1}a^{k - n + 1}))x^{2n - k}
\end{gather}
with $|A_1|, \ldots, |A_n|\gg |a_1|, \ldots, |a_n|$.  To match the coefficients, it is consistent to assume that $D$ is of order $A$ (with errors of order $a$) and $G$ is of order $Aa$ (with errors of order $a^2$).  Matching the coefficients with $0\leq k\leq n$ shows that
\begin{equation}
(A_1, \ldots, A_n) = \text{some permutation of $(d_1 + O(G/D), \ldots, d_n + O(G/D))$}.
\end{equation}
Then $A_1\cdots A_n = d_1\cdots d_n + O(D^{n-2}G)$, and further matching the coefficients with $n + 1 \linebreak[1] \leq \linebreak[1] k \linebreak[1] \leq \linebreak[1] 2n$ shows that
\begin{equation}
(-1)^{k-n}e_{k-n}(a_1, \ldots, a_n) = \frac{1}{d_1\cdots d_n}\sum_{|\alpha| = 2n - k} d_{\alpha(1)}\cdots d_{\alpha(2n - k)}\det(G(\alpha|\alpha)) + O(D^{n - k - 2}G^{k - n + 1})
\end{equation}
for $n + 1\leq k\leq 2n$, or more simply,
\begin{equation}
(-1)^k e_k(a_1, \ldots, a_n) = \frac{1}{d_1\cdots d_n}\sum_{|\alpha| = n - k} d_{\alpha(1)}\cdots d_{\alpha(n - k)}\det(G(\alpha|\alpha)) + O\left(\frac{G^{k + 1}}{D^{k + 2}}\right)
\end{equation}
for $1\leq k\leq n$.  We conclude that the $a_i$ are the roots of the polynomial
\begin{equation}
x^n + \sum_{k=1}^n \underbrace{\left[\frac{1}{d_1\cdots d_n}\sum_{|\alpha| = n - k} d_{\alpha(1)}\cdots d_{\alpha(n - k)}\det(G(\alpha|\alpha)) + O\left(\frac{G^{k + 1}}{D^{k + 2}}\right)\right]}_{O(G^k/D^k)}x^{n-k}.
\end{equation}
This is consistent with all the $a_i$ being of order $G/D$.

\section{Numerical Neutrino Oscillation Fit for \texorpdfstring{$n = 3$}{n = 3}}

In this section, we provide explicit numerical fits to the observed neutrino oscillation data and show that significant mixing between active and sterile neutrinos can be achieved for the case with three sterile neutrinos. We write
\begin{equation}
M = \left[\begin{array}{ccc|ccc}
0 & 0 & 0 & b_{11} & b_{12} & b_{13} \\
0 & 0 & 0 & b_{21} & b_{22} & b_{23}\\
0 & 0 & 0 & b_{31} & b_{32} & b_{33} \\ \hline
b_{11} & b_{21} & b_{31} & d_{11} & d_{12} & d_{13} \\
b_{12} & b_{22} & b_{32} & d_{12} & d_{22} & d_{23} \\
b_{13} & b_{23} & b_{33} & d_{13} & d_{23} & d_{33}
\end{array}\right].
\label{eq:mat}
\end{equation}
Our numerical method is based on a constrained minimization where five neutrino observables ($\Delta m_{21}^2$, $\Delta m_{31}^2$, $\sin^2\theta_{12}$, $\sin^2\theta_{13}$, $\sin^2\theta_{23}$) are forced to lie within their experimentally measured ranges. The values of the input parameters in Eq.\ \eqref{eq:mat} that yield fits to the oscillation parameters of Table \ref{nuTab3} are shown in Table \ref{tab:param1} and Table \ref{tab:param2} for {\tt Fit1} and {\tt Fit2}, respectively.

\clearpage

\vspace*{\fill}
{
\renewcommand{\arraystretch}{1.2}
\centering
\begin{tabular}{|c|c|c|c|}
\hline \hline
\textbf{Oscillation} & \textbf{$3\sigma$ Allowed Range} & \multicolumn{2}{c|}{\textbf{Model Fits}} \\
\cline{2-4}
\textbf{Parameters} & \textbf{{\tt NuFit5.2} \cite{Esteban:2020cvm}} & {\tt Fit1} & {\tt Fit2} \\ \hline \hline
$\Delta m_{21}^2$ ($10^{-5}$ eV$^2$) & 6.82 -- 8.03 & 7.41 & 7.55 \\ \hline
$\Delta m_{31}^2$ ($10^{-3}$ eV$^2$) & 2.428 -- 2.597 & 2.54 & 2.54 \\ \hline
$\sin^2{\theta_{12}}$ & 0.270 -- 0.341 & 0.3112 & 0.320 \\ \hline
% $\sin^2{\theta_{23}}$ (IH) & 0.419 - 0.617 & - & 0.589 \\
$\sin^2{\theta_{23}}$ & 0.406 -- 0.62 & 0.454 & 0.518 \\ \hline
% $\sin^2{\theta_{13}}$ (IH) & 0.02052 - 0.02428 & - & 0.0228 \\
$\sin^2{\theta_{13}}$ & 0.02029 -- 0.02391 & 0.0223 & 0.0213 \\ \hline
% $\delta_{CP}/^\circ$ (IH) & 193 - 352 & - & 318 \\
% $\delta_{CP}/^\circ$ & 120 - 369 & 269 & 277
%%%%%%%
\end{tabular}
\captionof{table}{$3\sigma$ allowed ranges for the neutrino oscillation parameters from a recent global fit \cite{Esteban:2020cvm}, along with the benchmark fits.} % For comparison, the $3\sigma$ allowed ranges for the oscillation parameters are also given.
\label{nuTab3}
\vspace{2cm}
%%%%%%%%%%%%%%%%%%%%%%%%%%%%%%%%%%%%%%%%%%%%%%%%%%%%%%
\centering
\begin{tabular}{|c|c|c|c|c|c|c|c|c|c|c|c|}
\hline \hline
& $b_{11}$ & $b_{12}$ & $b_{13}$ & $b_{21}$ & $b_{22}$ & $b_{23}$ & $b_{31}$ & $b_{32}$ & $b_{33}$ \\[3pt] \hline
%%%%%%%
& $8.348$ eV & $0.297$ GeV & $1.851$ eV & $3.35$ meV & $-1.325$ GeV & $0.572$ eV & $14.65$ eV & $1.453$ GeV & $1.30$ eV \\[3pt] \cline{2-10}
%%%%%%%%%%%%%
{\tt Fit1} & $d_{11}$ & $d_{12}$ & $d_{13}$ & $d_{22}$ & $d_{23}$ & $d_{33}$ & $|U_{24}|$ & $|U_{25}|$ & $|U_{26}|$ \\[3pt] \cline{2-10}
%%%%%%%
& $-6.306$ GeV & $1.132$ TeV & $-5.13$ TeV & 0 & $-1.897$ TeV & $d_{13}^2/d_{11}$ & $8.25 \times 10^{-5}$ & $8.25 \times 10^{-5}$ & 0 \\[3pt] \hline \hline
\end{tabular}
\captionof{table}{Values of parameters that generate {\tt Fit1} to the neutrino oscillation data, as shown in Table \ref{nuTab3}. Here, $(U_{24}, U_{25}, U_{26})$ are the mixing parameters between active and sterile neutrinos.}
\label{tab:param1}
\vspace{2cm}
%%%%%%%%%%%%%%%%%%%%%%%%%%%%%%%%%%%%%%%%%%%%%
\centering
\begin{tabular}{|c|c|c|c|c|c|c|c|c|c|c|c|}
\hline \hline
& $b_{11}$ & $b_{12}$ & $b_{13}$ & $b_{21}$ & $b_{22}$ & $b_{23}$ & $b_{31}$ & $b_{32}$ & $b_{33}$ \\[3pt] \hline
%%%%%%%
& $-0.338$ eV & $13.01$ GeV & $-0.15$ eV & $0.74$ meV & $-141$ GeV & $0.176$ eV & $-4.0$ eV & $-66.82$ GeV & $0.36$ eV \\[3pt] \cline{2-10}
%%%%%%%%%%%%%
{\tt Fit2} & $d_{11}$ & $d_{12}$ & $d_{13}$ & $d_{22}$ & $d_{23}$ & $d_{33}$ & $|U_{24}|$ & $|U_{25}|$ & $|U_{26}|$ \\[3pt] \cline{2-10}
%%%%%%%
& $-0.881$ TeV & $22.92$ GeV & 0 & $1.47$ TeV & $-2.293$ TeV & 0 & $0.001$ & $0.0495$ & $0.036$ \\[3pt] \hline \hline
\end{tabular}
\captionof{table}{Values of parameters that generate {\tt Fit2} to the neutrino oscillation data, as shown in Table \ref{nuTab3}. Here, $(U_{24}, U_{25}, U_{26})$ are the mixing parameters between active and sterile neutrinos.}
\label{tab:param2}
}
\vspace*{\fill}

\clearpage

\end{document}